\documentclass[12pt, letterpaper] {article}
\newcommand{\be}{\begin{equation}}
\newcommand{\ee}{\end{equation}}
\usepackage{amsmath,amssymb,array,calc,amsfonts,latexsym,color}   
\usepackage{graphicx}  
\usepackage{tensind}  
\tensordelimiter{?}

\usepackage{multirow}

\long\def\symbolfootnote[#1]#2{\begingroup%
\def\thefootnote{\fnsymbol{footnote}}\footnote[#1]{#2}\endgroup}
\numberwithin{equation}{section}
\usepackage{hyperref}
\begin{document}

\begin{titlepage}
\vskip -.8cm

\rightline{\small{\tt MCTP-09-20}}

\vskip .7 cm

\centerline{\bf \Large Tensions and L\"uscher Terms for (2+1)-dimensional $k$-strings}

\vskip .5 cm

\centerline{\bf \Large from Holographic Models }
\vskip 1cm

\centerline{\large Christopher A. Doran${}^1$\symbolfootnote[1]{christopher-doran@uiowa.edu}, Leopoldo A. Pando Zayas${}^2$\symbolfootnote[2]{lpandoz@umich.edu},}
\centerline{\large Vincent G. J. Rodgers${}^1$\symbolfootnote[3]{vincent-rodgers@uiowa.edu} and Kory Stiffler${}^1$\symbolfootnote[4]{kstiffle@gmail.com}}

\vskip .4cm

\centerline{\it ${}^1$Department of Physics and Astronomy}
\centerline{ \it  The University of Iowa}
\centerline{\it Iowa City, IA 52242}

\vskip .3cm

\centerline{\it ${}^2$Michigan Center for Theoretical
Physics}
\centerline{ \it Randall Laboratory of Physics, The University of
Michigan}
\centerline{\it Ann Arbor, MI 48109-1040}

\vspace{.5cm}


\begin{abstract}
\baselineskip15pt
The leading term for the energy of a bound state of $k$-quarks and $k$-antiquarks is proportional to its separation $L$. These $k$-string configurations have a L\"uscher term associated with their quantum fluctuations which is typically a 1/$L$ correction to the energy. We review the status of tensions and L\"uscher terms in the context of lattice gauge theory, Hamiltonian methods, and gauge/gravity correspondence. Furthermore we explore how different representations of the $k$-string manifest themselves in the gauge/gravity duality.  We calculate the L\"uscher term for a strongly coupled $SU(N)$ gauge theory in $(2+1)$ dimensions using the gauge/gravity correspondence. Namely, we compute one-loop corrections to a probe  D4-brane embedded in the Cveti$\check{c}$, Gibbons, L\"u, and Pope supergravity background.  We investigate quantum fluctuations of both the bosonic and the fermionic sectors.
\end{abstract}

\end{titlepage}

\section{Introduction}
Since the emanation of Quantum Chromodynamics as the theory of the strong nuclear force, serious theoretical and predictive challenges appeared due to the strongly coupled nature of the theory at nuclear energies.  What has emerged from these challenges is a technical and computational acumen that has advanced our understanding of other theories as well.   Lattice gauge theory, advances in the Hamiltonian formulation of gauge theories, and string theories (including the gauge/gravity correspondence) are three examples of these powerful theoretical tools that have emerged from this quest.
At present the ability for theorists to match experimental data in the strongly coupled regime is very far from the success achieved in the perturbative regime.  For this reason, theorist sometimes resort to comparing theories among themselves for states that are suitably accessible.  For example the mass gap of the pure Yang Mills sector has been observed by the lattice community for some time and through string theory as well.   For string theory, Polyakov gave a description of the mass gap by including the extrinsic curvature to Nambu-Goto action \cite{Polyakov:1986cs}, where one can show that asymptotic freedom implies a mass gap. The question of the mass gap for $2+1$ dimensions was also addressed using the Hamiltonian formulation in  \cite{Karabali:1997wk,Karabali:2000gy}. Through this one sees that certain string theories and gauge theories belong to the same universality class.  Suitable configurations that are calculationally accessible to all of these methods give theorists a standard for comparison.  The $k$-string epitomizes such configurations. In this note, we examine the gauge/gravity dual for a $2+1$ dimensional $k$-string configuration using the Cveti$\check{c}$, Gibbons, L\"u, and Pope \cite{Cvetic:2001ma} (CGLP) supergravity background.  Because this background is dual to a $2+1$ dimensional gauge theory, this configuration lends itself well to comparisons of both lattice gauge theories and Hamiltonian approaches to gauge theories.

Here we give a brief overview of $k$-strings and an outline of this note. For a more complete review of $k$-strings, see~\cite{Shifman:2005eb}.
$k$-strings are a configuration of $SU(N)$ color sources which result from $k$ color sources in the fundamental representation stretched a large distance $L$ from $k$ anti-color sources. The gauge/gravity correspondence has been used to explore the relationship between certain configurations of low energy supergravity backgrounds and low energy $k$-strings  ~\cite{hk,Herzog:2002ss,Ridgway:2007vh,tye,Zayas:2008hw}.  Many of these supergravity calculations have been done for $D3$-branes, which are dual to four dimensional Yang-Mills theories.  On the gauge theory and lattice gauge theory side of the correspondence, much of the focus has been in three dimensions\cite{Ambjorn:1984mb,Ambjorn:1984yu,Ambjorn:1984me,Karabali:1998yq,Teper:1998te,Bringoltz:2006zg,Bringoltz:2008nd}.  Therefore the best test of the gauge/gravity correspondence through $k$-string configurations is to find supergravity backgrounds which are dual to $3d$ gauge theories.  We will review in this paper  the Cveti$\check{c}$, Gibbons, L\"u, and Pope (CGLP) type IIA solution which is one such supergravity background \cite{Cvetic:2001ma}.

From the gauge theory side of things, the classical energy of strongly coupled $k$-strings is found to follow either a sine law or a Casimir law:
\begin{align}
   E_k \propto T_k L~~~\mbox{$k$-string energy} \\
   \label{eq:sinelaw}
   T_k \propto N \sin\frac{k \pi}{N}~~~\mbox{sine law} \\
   ~~~~~~~~~~~~~\mbox{or}            \nonumber\\
   \label{eq:Casimirlaw}
   T_k \propto k(N-k)~~~\mbox{Casimir law}.
\end{align}
The precise form of the tension appears to be inconclusive \cite{Ambjorn:1984mb,Ambjorn:1984yu,Ambjorn:1984me,Karabali:1998yq,Teper:1998te,Bringoltz:2006zg,Bringoltz:2008nd,Armoni:2003ji,Armoni:2003nz}.

In this paper, we will investigate specifically the gauge theory dual to $N$ stacked $D2$-branes, and stacked fractional $D2$-branes sourcing the CGLP type $IIA$ supergravity background~\cite{Cvetic:2001ma}.  In this background, the low energy spectrum has been found to have a slightly lower tension than both the Casimir and sine laws ~\cite{Herzog:2002ss}.

We probe this calculation further, by calculating the one loop quantum correction to the low energy classical solution of the supergravity. Using vanishing boundary conditions on the probe $D$-brane, we find the energy correction is dual to an $SU(N)$ L\"uscher term with vanishing boundary conditions (i.e., the $k$-string ends on immovable color sources),
\begin{align}
  V_{\mbox{L\"uscher}} &= -\frac{\pi}{6 L}.
\end{align}
\noindent 
Our results constitute a prediction for the L\"uscher term for the $k$-string.
When comparing this result with the  L\"uscher term  coming from the fundamental string found in ~\cite{Luscher:1980fr,Luscher:1980ac,Lucini:2001nv}, viz.  $-\pi/24 L$, one should be mindful to the fact that the $k$-string configuration in the gauge/gravity correspondence has its origins in a large $N$ setting where $k/N$ is held fixed.  In the same vein, naively setting $k=1$ in the gauge/gravity correspondence may not recover the fundamental string.

\subsection{$k$-String Ground State}
The $k$-string tension provides the first level of comparison for various theories.
In the context of ${\cal N}=2$ supersymmetric theories,
Douglas and Shenker \cite{Douglas:1995nw}, examined the $N$-extended monopole condensation model of Seiberg and Witten for $SU(2)$ \cite{Seiberg:1994aj} and  found a spectrum of string tension that obeys a sine law, $T_k \sim N \sin{(\frac{\pi k}{N})}$. In a precursor to AdS/CFT, Hanany, Strassler and Zaffaroni were able to reproduce this spectrum of meson by using an M-theory fivebrane approach to QCD (MQCD) \cite{Hanany:1997hr}.  The lattice community has actively studied the $k$-strings tension for some time \cite{Lucini:2004my,Lucini:2001nv,DelDebbio:2001kz}. One of the issues that arises is whether these configurations exhibit a Casimir-like  scaling \[ T_k\approx k(1-k/N) \] for large N or a sine law where \[ T_k \propto N \sin{(\frac{\pi k}{N})}.\] Both of these behaviors respect the $N$-ality but on the one hand, the $1/N$ expansion of QCD agrees with the sine law scaling  \cite{Armoni:2003ji,Armoni:2003nz}, while on the other hand lattice calculations of \cite{Bringoltz:2008nd} favor Casimir scaling ($1/N$).

\subsubsection{The Wealth of Information in $2+1$ Dimensions}
Because the analysis is tractable, examining the $k$-string configurations in $2+1$ dimensions gives a wealth of information. In \cite{Karabali:1998yq} Karabali, Kim and Nair predict that the string tensions in $SU(N)$ gauge theories should be
proportional to the quadratic Casimir of the representation of the flux.  Based
on the idea of effective dimensional reduction driven by a highly disordered vacuum, it was conjectured a long time ago \cite{Ambjorn:1984mb,Ambjorn:1984yu,Ambjorn:1984me} that this might also hold in $D = 2+1$ and $D = 3+1$. There is some additional evidence for this hypothesis from calculations of the potential between charges in various representations of $SU(3)$.

For a given $k$ the smallest Casimir arises for the totally antisymmetric representation, and this should therefore provide the ground state k-string tension:
\be
\frac{\sigma_k}{\sigma_f}=\frac{k(N-k)}{N - 1}.
\label{eqn9}
\ee
This is the part of the 'Casimir Scaling' hypothesis that we shall be mainly testing in this paper. For this purpose it is useful to have an alternative conjecture that possesses the correct general properties. A convenient and well-known example is provided by the trigonometric form
\be
\frac{\sigma_k}{\sigma_f}=\frac{\sin \frac{k\pi}{N}}{\sin\frac{\pi}{N}},
\ee
that was originally suggested on the basis of an M-theory approach to QCD \cite{Hanany:1997hr} and has appeared in the context of the gauge/gravity correspondence. In fact the full prediction of Karabali et al. for $\sigma_k$ is more specific than Eq.(\ref{eqn9}), since it also predicts a value for $\sigma_f$ in terms of $g^2$, and including this gives:
\be
\sigma_R=e^4 \frac{C_A \, C_R}{4\pi},
\ee
where $C_A$ is the quadratic Casimir invariant for the adjoint representation defined as $f^{acd}f^{bcd}=C_A\delta^{ab}$. This may be written as:
\be
\frac{\sigma_k}{(g^2N)^2}=\frac{1}{8\pi} \frac{k(N -k)(N + 1)}{N^2}.
\label{eqn11}
\ee
Recent improvements to the Hamiltonian prediction of Karabali et al. \cite{Karabali:2009rg} have moved the value of $\sigma_f$ to within $-.3\%$ to $-2.8\%$ of the lattice value.

\subsubsection{Representations for $k$-strings}

In \cite{Gomis:2006sb,Gomis:2006im}, Gomis and Passerini studied various representations of Wilson loops in ${\cal N}=4$ supersymmetric Yang-Mills as related to particular representations of the gauge group.  Inspired by this, we postulate that different supergravity backgrounds in the gauge/gravity correspondence can be identified with symmetric and antisymmetric representations of the $k$-string.  For the case of $2+1$, both the Cveti$\check{c}$, Gibbons, L\"u, and Pope (CGLP) \cite{Cvetic:2001ma} and the Maldacena-Nastase (M-Na) \cite{Maldacena:2001pb} supergravity backgrounds are relevant.  We identify the following configurations with the group representations:
\begin{itemize}
\item the probe D4 brane with world volume flux wrapping a three sphere in the CGLP background = antisymmetric representation;
\item
the probe D3 brane with world volume flux wrapping a two sphere in the M-Na background = symmetric representation.
\end{itemize}
We postulate that in the holographic context, at least in $2+1$ dimensional theories, it seems that both types of formulas are possible: Casimir for CGLP and sine law for M-Na. Moreover, this suggests that the antisymmetric representations for the holographic case (CGLP) are ``Casimir-law" while the symmetric representations (M-Na) are ``sine-law".
\eject
\begin{table}[h!]
\centering
\begin{tabular}{|c|c|c|c|c|c|c|}
\multicolumn{7}{c}{$T_k/T_f$ from Various Methods} \\
\multicolumn{7}{c}{S=symmetric, A=antisymmetric, M=mixed, *=antisymmetric} \\
\hline
$Group$ & $k$ &  CGLP & MNa(Sine) & Casimir & lattice & Karabali-Nair  \\
\hline\hline
\multirow{2}{*}{$SU(4)$} & \multirow{2}{*}{2}  & \multirow{2}{*}{1.310} & \multirow{2}{*}{1.414}  & \multirow{2}{*}{1.333} &  1.353(A) & 1.332(A)   \\
&&&&& 2.139(S) & 2.400(S)\\
\hline\hline
$SU(5)$ & 2 &  1.466 & 1.618 & 1.5 & 1.528* & 1.529*  \\
\hline\hline
\multirow{5}{*}{$SU(6)$} & \multirow{2}{*}{2} & \multirow{2}{*}{1.562} & \multirow{2}{*}{1.732} & \multirow{2}{*}{1.6} & 1.617(A) & 1.601(A) \\
\cline{6-7}
&&&&& 2.190(S) & 2.286(S)\\
\cline{2-7}
& \multirow{3}{*}{3} & \multirow{3}{*}{1.744} & \multirow{3}{*}{2.0} & \multirow{3}{*}{1.8} & 1.808(A) & 1.800(A) \\
\cline{6-7}
&&&&&3.721(S) & 3.859(S) \\
\cline{6-7}
&&&&&2.710(M) & 2.830(M)\\
\hline\hline
\multirow{3}{*}{$SU(8)$} & 2 & 1.674 & 1.848 & 1.714 & 1.752* & 1.741*\\
\cline{2-7}
&3 & 2.060 & 2.414 & 2.143 & 2.174* & 2.177*\\
\cline{2-7}
&4 & 2.194 & 2.613 & 2.286 & 2.366* & 2.322*\\
\hline
\end{tabular}

\caption{Comparison of $k$-string tensions from various methods.  The values quoted are $T_k/T_f$, where $T_k$ is the $k$-string tension, and $T_f$ is the fundamental string tension, i.e., $k=1$.  The CGLP tension is calculated from the transcendental Eqs.(\ref{eq:Hmin},\ref{eq:mincon}); MNa(Sine) from Eq.(\ref{eq:MNaTension}); Casimir from Eq.(\ref{eq:Casimirlaw}). The A, S, and M data are calculated directly from~\cite{Karabali:2007mr}, the * data is quoted directly from~\cite{Bringoltz:2008nd}.}
\label{tab:comparetensions}
\end{table}

Table~\ref{tab:comparetensions} shows this solution for supergravity backgrounds  with lattice calculations by Bringholtz and Teper and Yang-Mills Hamiltonian calculations by Karabali et al. All calculations are in $2+1$ dimensions where the  quarks are in the anti-symmetric, symmetric, or mixed representations.  Table~\ref{tab:comparetensions} clearly shows that the CGLP tensions are lower than the Casimir law, lattice data and Karabali et al. calculations, and that the M-Na tensions are all higher. The full Table in $2+1$, including all possible supergravity solutions with their corresponding brane embeddings,  is not yet known.  For example,other configurations to be considered are the D6 branes in the CGLP background which would correspond to a symmetric state, D5 brane in the MNa background~\cite{Maldacena:2001pb}(antisymmetric), and the Witten 2+1 QCD (nonextremal D3-brane compactified on a circle).  As far as the explicit values of the tensions, our suggestion for which supergravity configuration/$k$-string representation correspondence is not well represented by the table. Both the CGLP and the M-Na tensions are more consistent with the antisymmetric representations.  Further studies are needed to validate this issue and determine how these geometries manifest the $k$-string tensor representations.  The question still remains as to where the mixed representations would fit into the supergravity scenarios.

\subsection{Beyond the ground state: The L\"uscher term}
A detailed study of the flux tube between a quark and an anti-quark is an important window into the physics of confinement. A penetrating approach in this study is to consider the effective action for such string as an expansion in derivative terms. L\"uscher and Weisz considered all terms allowed by symmetries and built an effective theory for the excitations \cite{Luscher:2004ib} and then went on to study the influence of the various terms on observables.  Aharony and Katzburn were able to take this up to six derivative couplings and show that the full spectrum of the theory only depends on two free parameters \cite{Aharony:2009gg}, the string tension $\sigma$  and a regularization dependent mass $\mu$. The quark-antiquark potential can be computed and shown to be
\be\label{eq:LuscherTerm}
V (L) = \sigma L +\mu +\frac{\gamma}{L}+ {\cal O}(1/L^2), \qquad \gamma=-\frac{\pi}{24}(d-2),
\ee
for  large distances $L$, where $d$ is the dimension of space-time. The $1/L$
correction in this formula is a quantum effect and can be used to determine a universality class for a large family of strings \cite{Luscher:1980ac}.
One remarkable property of the above expansion is that it sets in for relatively small values of $L$, for example it was argued in \cite{Luscher:2004ib} that for values around $0.5{\rm fm}$ the above expansion is already valid.

Furthermore, there is an exponential reduction of the statistical errors in the lattice calculations of L\"uscher and Weisz \cite{Luscher:2002qv}
 which  allows them to compute  the quark-antiquark potential with high accuracy.  Their results are completely consistent with theoretical model within a $0.04\%$ and $0.12\%$ depending on the quark-antiquark separation.

We wish to compare the L\"usher term in Eq.(\ref{eq:LuscherTerm}) to that calculated from holographic models.  The fundamental string is easy to find in most holographic models. It corresponds to a static fundamental string whose main contribution to the energy comes form the IR region of the dual supergravity background. Analyzing the massless modes of such fundamental strings in various confining backgrounds is rather simple since it is equivalent to counting symmetries. Since only those massless modes contribute to the L\"uscher term we conclude that:

\begin{eqnarray}
\gamma^{fund}_{3d}&=& -\frac{\pi}{24}, {\rm (CGLP~\cite{Cvetic:2001ma}, M-Na~\cite{Maldacena:2001pb}\,\, backgrounds)}, \nonumber \\
\gamma^{fund}_{4d}&=& -\frac{\pi}{12}, {\rm  \mbox{(KS~\cite{ks}, M-N\'u\~nez~\cite{Maldacena:2000yy} backgrounds)}}.
\end{eqnarray}
As emphasized before, the above results come only from the massless modes. Namely  in 3-dimensional models we have only one massless mode viz: $X^i$, $i=2$.  Similarly, in 4-dimensional models, there are two massless modes: $X^{i}$ with $i=2,3$, while all other modes are massive, including the fermions.

Another very interesting configuration in confining theories is the one formed by $k$ quarks separated a distance $L$ by $k$ anti-quarks. This configuration, known as the $k$-string, also contains an intricate structure of excitations. We have studied such configurations at one-loop level in various supergravity backgrounds. We find that for both the KS and CGLP backgrounds, the four-dimensional\footnote{Here we correct a factor of a half missing in~\cite{Zayas:2008hw}.} and three-dimensional theories have the same L\"uscher term:
\begin{equation}
\gamma^{k-string}_{4d} = \gamma^{k-string}_{3d} = -\frac{\pi}{6}.
\end{equation}

There are four massless modes for each.  Let us explain: from the holographic point of view the difference comes from the fact that we have the standard massless Goldstone mode for the fluctuations in the transverse direction to the corresponding brane configuration, whereas the extra massless modes with respect to those of the fundamental string arise from the gauge field that is necessary to include holographically to get a representation characterized by $k$. In the case of the KS background our computation shows that there are two extra massless degrees of freedom coming from the fluctuations of the gauge field \cite{Zayas:2008hw}. We verify later in this paper that in the three-dimensional case, the gauge field contributes three massless degrees of freedom to the L\"uscher term.

\section{$k$-strings from type II Supergravity Duals}
\subsection{The Basic Idea}
In this section, we give a brief sketch of how to calculate low energy $k$-strings from a dual supergravity theory.  The $k$-string is thought of, from the supergravity dual, as a \emph{probe} Dp-brane with electromagnetic charge in its world volume

\begin{align}
    F = dA = F_{tx}~dt \wedge dx + F_{\theta\phi}~d\theta \wedge d\phi,
\end{align}

\noindent embedded in a supergravity background.

This probe Dp-brane will \emph{wrap}, or be tangent to, $p+1$ out of the 10 bosonic supergravity coordinates, $X^{\mu}$.  The remaining $9-p$ supergravity coordinates will act as \emph{scalar} fields, with dynamics on the probe Dp-brane.  These scalar fields will enter into the action for the probe Dp-brane through the dilaton, $\Phi$, and the pullbacks of the other bosonic supergravity sources
\begin{align}\label{eq:pforms}
 F_{n+1}=dC_n,~~~H_3=dB_2.
\end{align}

In addition, we will have fermionic fields with dynamics on our probe Dp-brane as well.  The action we use for the probe Dp-brane, parametrized by $\zeta$, is
\begin{align}\label{eq:SDp}
    S_{Dp} &= -\mu_p \int d^{p+1}\zeta~ e^{-\Phi} \sqrt{-\det{(g_{ab} + \mathcal{F}_{ab})}} + \mu_p \int e^{\mathcal{F}} \wedge \sum_q C_q + S_f \\
    \mathcal{F} &= B_2 + 2\pi\alpha'F, ~~~\mu_p^{-1} = (2\pi)^p \alpha'^{(p+1)/2}.
\end{align}

\noindent where all of the fermionic fields are contained in $S_f$.  The induced metric, $g_{ab}$, is the pullback of the 10 dimensional supergravity metric $G_{\mu\nu}$
\begin{align}
   g_{ab} &= \frac{\partial X^{\mu}}{\partial \zeta^a}\frac{\partial X^{\nu}}{\partial \zeta^b} G_{\mu\nu},
\end{align}

\noindent The generalization of this formula is how the n-forms listed in Eq.(\ref{eq:pforms}) are pulled back to the $Dp$-brane.

For the low energy solution, we follow~\cite{Herzog:2002ss} and set the fermions all to zero, $S_f = 0$, consider bosonic fields constant on the Dp-brane coordinates, $X^{\mu} = X^{\mu}_0$ = constant, and integrate out the spatial Dp-brane coordinates from the action, Eq.(\ref{eq:SDp}), leaving us with
\begin{align}
   S &= \int dt \mathcal{L}(A_a,\dot{A}_a,X^{\mu}_0)
\end{align}
This particular solution leaves only the gauge field on the probe Dp-brane, $A_a$ with any dynamics with which to calculate the low energy Hamiltonian, which is proportional to the $k$-string tension, $T_k$:
\begin{align}
   H &= \frac{\partial \mathcal{L}}{\partial \dot{A}_a}\dot{A}_a - \mathcal{L} = L T_k
\end{align}

\subsection{Review of the CGLP Background}
Let us first review the CGLP type IIA supergravity solution.  More details can be found in ~\cite{Cvetic:2001ma,Herzog:2002ss,Cvetic:2000mh}.  The CGLP background is a solution with $N$ coincident D2-branes

In the string frame, the CGLP solution with these sources is found to be
\begin{align}\label{eq:CGLP}
  ds_{10}^2 &= G_{\mu\nu}dX^{\mu}dX^{\nu} = H^{-1/2}dx^{\alpha}dx^{\beta}\eta_{\alpha\beta} + H^{1/2}ds_7^2,\\
  \label{eq:dilaton}
  e^{\Phi} &= g_s H^{1/4}
\end{align}

\noindent where $\eta_{\alpha\beta}$ is $\mathbb{R}^{1,2}$, and
\begin{align}\label{eq:metric7d}
  ds_7^2 &= l^2[h^2 dr^2 + a^2(D\mu^i)^2 + b^2 d\Omega_4^2], \\
  \label{eq:identities}
  X_2 &\equiv \frac{1}{2}\epsilon_{ijk}\mu^iD\mu^i\wedge D\mu^k,~~~J_2 \equiv \mu^i J^i,~~~ X_3 \equiv dX_2 = dJ_2
\end{align}

\noindent In the above, $l$, $m$, and $g_s$ are constants, and $a,b,h,u_i$ and $H$ are functions of $r$.
\begin{align}\label{eq:hab}
  h^2 &= (1 - r^{-4})^{-1},~~~a^2 = \frac{1}{4}r^2(1 - r^{-4}),~~~b^2 = \frac{1}{2}r^2 \\
  u_1 &= r^{-4} + P(r)r^{-5}(r^4 -1)^{-1/2},~~~~u_2 = -\frac{1}{2}(r^4 - 1)^{-1} + P(r)r^{-1}(r^4 -1)^{-3/2}, \nonumber\\
  u_3 &= \frac{1}{4}r^{-4}(r^4-1)^{-1} - \frac{3r^4 -1}{4r^5(r^4 - 1)^{3/2}} P(r) \\
  P(r) &= \int_1^r \frac{d\rho}{\sqrt{\rho^4 - 1}} \\
  H(r) &= \frac{m^2}{2l^6}\int_r^\infty \rho (2 u_2(\rho) u_3(\rho) - 3 u_3(\rho))d\rho.
\end{align}

\noindent The parameter $l$ is similar to $\epsilon$ in the deformed conifold~\cite{Herzog:2002ss,candelas,ks} and $g_s$ is the string coupling constant.

The differential element $D\mu^i$ is
\begin{align}
  D\mu^i = d\mu^i + \epsilon_{ijk}A^j \mu^k
\end{align}

\noindent where the $\mu^i$ are coordinates on a unitless $\mathbb{R}^3$ constrained  to a unit $S^2$ surface, $\mu^i\mu^i = 1$.

The fluxes associated with the solution are:

\begin{align}\label{eq:H3}
H_3 &= \frac{m}{l}a^2 u_1 h dr \wedge X_2 + \frac{m}{l}b^2 u_2 h dr \wedge J_2 + \frac{m}{l}a b^2 u_3X_3,~~~C_1 = 0 \\
\label{eq:F4}
F_4 &= g_s^{-1}d^3x \wedge dH^{-1} + m g_s^{-1}G_4, \\
  G_4 &= a b^2 u_3~\epsilon_{ijk}~\mu^i~h dr \wedge D\mu^j \wedge J^k + a^2b^2u_2 X_2 \wedge J_2 + \frac{1}{2}b^4 u_1 J_2 \wedge J_2
\end{align}

\noindent where $d^3x \equiv dx^0 \wedge dx^1 \wedge dx^2$.

The $A^i$ are $SU(2)$ Yang-Mills instanton one forms living on the $S^4$
\begin{align}
   A^i &= A^i_{\alpha}d\Omega_4^{\alpha},~~~d\Omega_4^{\alpha} = (d\psi,d\chi,d\theta,d\phi).
\end{align}
\noindent  and compose an anti-symmetric $SU(2)$ Yang-Mills two form, $J^i$,
\begin{align}
  J^i &= dA^i + \frac{1}{2}\epsilon_{ijk}A^j \wedge A^k,
\end{align}

\noindent which satisfies the algebra of the unit quaternions,
\begin{align}
   \hat{g}^{\gamma\rho}J^i_{\alpha\gamma}J^j_{\rho\beta}&= -\delta_{ij}\hat{g}_{\alpha\beta} + \epsilon_{ijk}J^k_{\alpha\beta},
\end{align}

\noindent where $\hat{g}_{\alpha\beta}$ is the metric for an $S^4$:
\begin{align}
   d\Omega^2_4 \equiv \hat{g}_{\alpha\beta}d\Omega^{\alpha}_4 d\Omega^{\beta}_4 =  d\psi + \sin^2\psi(d\chi^2 +\sin^2\chi(d\theta^2 + \sin^2\theta d\phi^2)).
\end{align}

\noindent As in~\cite{Cvetic:2001ma,Herzog:2002ss}, we select the $J^i$ to be
\begin{align}
   J^1 &= -\sin\psi~d\psi\wedge d\chi - \sin^2\psi~\sin^2\chi~\sin\theta d\theta\wedge d\phi \\
   J^2 &= -\sin\psi~\sin\chi~d\psi\wedge d\theta - \sin^2\psi~\sin\chi~\sin\theta d\phi\wedge d\chi \\
   J^3 &= -\sin\psi~\sin\chi~\sin\theta d\psi\wedge d\phi - \sin^2\psi~\sin\chi d\chi\wedge d\theta
\end{align}

\noindent and we select the gauge where the solution for $A^i$ is
\begin{align}\label{eq:Asu2}
  A^1 &= \cos\psi d\chi + \cos\theta d\phi \nonumber\\
  A^2 &= \cos\psi~\sin\chi d\theta -\cos\chi~\sin\theta d\phi \nonumber\\
  A^3 &= \cos\psi~\sin\chi~\sin\theta d\phi + \cos\chi d\theta
\end{align}

We use the relations given in Eq.(\ref{eq:pforms}) to calculate $B_2$ and $C_3$.
In calculating $B_2$, the identities in Eq.(\ref{eq:identities}) are very helpful.  Using these, we solve for $B_2$, up to a total derivative, to be
\begin{align}
  lB_2 &= m\left(\int_1^r f_1(u)du\right)X_2 + m\left(\int_1^r f_2(r)du\right)J_2 \nonumber\\
       f_1(u) &= a^2(u) u_1(u)h(u) ,~~~f_2(u) = b^2(u)u_2(r)h(u).
\end{align}

\noindent Notice that this vanishes when $r=1$.  This will be important for our ensuing calculations as it is where we will position our probe D4-brane.
We choose the following solution for $C_3$:
\begin{align}
   C_3 &= -\frac{x^2 dr\wedge dx^0 \wedge dx^1}{g_s H(r)^2} + \frac{3m}{8g_s}\xi(\psi)d\Omega_3 + \frac{m}{2 g_s}u_2(r)b(r)^2a(r)^2\epsilon_{ijk}\mu^iD\mu^j \wedge J^k\\
  \xi(\psi) &= \int_0^{\psi} \sin^3u~du,~~~d\Omega_3 \equiv \sin^2\chi~\sin\theta~d\chi\wedge d\theta \wedge d\phi,
\end{align}

\noindent which is the same as that chosen in~\cite{Herzog:2002ss}, up to a total derivative.

The constant $m$ is proportional to the number $N$ of stacked fractional D2-branes that the background describes, which by the gauge/gravity correspondence, it is also the number $N$ of colors for the dual supersymmetric $SU(N)$ gauge theory.  We can calculate the proportionality constant by using the Dirac quantization condition~\cite{Herzog:2002ss}
\begin{align}
   \int_{S^4} F_4 &= 8\pi^3 \alpha'^{3/2} N,
\end{align}

\noindent and the $r \to 1$ limiting behavior for $G_4$
\begin{align}
   \int G_4 &\to \frac{3}{8}\int d\Omega_4.
\end{align}

\noindent Using these two conditions Eq.(\ref{eq:F4}) gives $m=8\pi\alpha'^{3/2} g_s N$.

Also, we must mention that there is another CGLP solution. To acquire the other solution, $d\Omega_4^2$ can be substituted for a metric over $\mathbb{CP}^2$. In this paper we will work only with the $4$-sphere, $d\Omega_4^2$.

\subsection{A Coordinate Transformation}
There is a singularity in the metric, Eq.(\ref{eq:CGLP}), at $r=1$, because the function $h(r)$, Eq.(~\ref{eq:hab}), blows up here.  This can be remedied by applying the coordinate transformation
\begin{align}\label{eq:coordinatetransformation}
  \tau = \sqrt{r -1 }.
\end{align}

Now the point $r=1$ is described by $\tau=0$ and the metric becomes
\begin{align}
  ds_7^2 &= l^2\left(f(\tau)^2d\tau^2 + a^2(D\mu^i)^2 + b^2 d\Omega_4^2 \right) \nonumber\\
   f(\tau) &= \frac{2 \tau}{\sqrt{1 - (1 + \tau^2)^{-4}}}.
\end{align}

\noindent Notice that $f(\tau)$ is finite as $\tau \to 0$.  Here are the expansions of all the aforementioned relevant functions in the $\tau \to 0$ limit.
\begin{align}
a(r(\tau)) &= \tau + O(\tau^3),~~~ b(r(\tau)) = \frac{1}{\sqrt{2}}(1 + \tau^2), \nonumber\\
u_1(r(\tau)) &= \frac{3}{2} - 7\tau^2 + O(\tau^4), ~~~u_2(r(\tau)) = -\frac{1}{4} + \frac{7}{10}\tau^2 + O(\tau^4),\nonumber\\
u_3(r(\tau)) &= -\frac{1}{4} + \frac{7}{5}\tau^2 + O(\tau^4),~~~f(\tau) = 1 + \frac{5}{4} \tau^2 + O(\tau^4),\nonumber\\
H(r(\tau)) &= H_0 - H_2\tau^2 + O(\tau^4)),~~~H_0 = \frac{m^2}{l^6}I_0,~~~H_2 = \frac{m^2}{l^6}\frac{7}{16}, \nonumber\\
I_0 &\equiv \int_1^\infty \rho (2 u_2(\rho) u_3(\rho) - 3 u_3(\rho))d\rho \approx 0.10693\dots
\end{align}

One can use these expansions to show that $B_2$ and $C_3$ become, under the coordinate transformations Eq.(\ref{eq:coordinatetransformation}),
\begin{align}\label{eq:B2tau}
  B_2 &= -\frac{m}{8l}\tau J_2 + O(\tau^3) \\
  \label{eq:C3tau}
  C_3 &= -\frac{2 x^2 \tau d\tau\wedge dx^0\wedge dx^1}{g_s H(r(\tau))^2} + \frac{3m}{8 g_s}\left( \xi(\psi)d\Omega_3 - \frac{1}{6}\tau^2 \epsilon_{ijk}\mu^iD\mu^j\wedge J^k\right) + O(\tau^4)
\end{align}

\subsection{Calculation of the $k$-string tension}
We now outline the calculation of the k-string tension already presented by~\cite{Herzog:2002ss}.
We label the world volume coordinates of the probe D4-brane as
\begin{align}
   \zeta^a &= (t, x, \chi,\theta,\phi),
\end{align}

\noindent where we are using the static gauge, and have fixed five of the bosonic supergravity coordinates to these D4-brane coordinates:
\begin{align}\label{eq:staticgauge}
   X^{\mu} &= (t, x, x^2, \psi, \chi, \theta, \phi, \mu^1,\mu^2,\mu^3, \tau).
\end{align}

\noindent These 11 coordinates are really 10 independent bosonic coordinates, as the $\mu$'s are constrained to $(\mu^i)^2 = 1$.

We have then, the low energy action for the probe D4-brane, embedded in the CGLP background:
\begin{align}\label{eq:SD4}
   S^{(b)} &= -\mu_4 \int d^5\zeta~ e^{-\Phi} \sqrt{-\det{(g_{ab} + \mathcal{F}_{ab})}} + \mu_4\int    \mathcal{F} \wedge C_3.
\end{align}

\noindent  Here, $B_2$ and $C_3$ have been pulled back to the D4-brane, the same way as the induced metric:
\begin{align}
  g_{ab} = \frac{\partial X^{\mu}}{\partial\zeta^a} \frac{\partial X^{\nu}}{\partial\zeta^b} G_{\mu\nu}.
\end{align}

We also turn on a $U(1)$ gauge flux on the D4-brane
\begin{align}
  2 \pi \alpha' H_0^{1/2}  F = E~dt \wedge dx
\end{align}

\noindent Since our D4-brane is placed at $\tau = 0$, where $B_2 = 0$, $\mathcal{F}$ becomes simply
\begin{align}\label{eq:curlyF}
  H_0^{1/2} \mathcal{F} = H_0^{1/2}(2\pi\alpha' F) = E~dt \wedge dx
\end{align}

We examine the action Eq.(\ref{eq:SD4}) in static gauge, Eq.(\ref{eq:staticgauge}), and at the classical solution
\begin{align}\label{eq:Solution}
  \tau &= x^2 = \mu^1 = \mu^2 = 0 ~~~\mu^3 = 1 ~~~\psi \equiv \psi_0 .
\end{align}

Here we calculate the induced metric to be
\begin{align}\label{eq:ginduced0}
   ds_0^2 &= g_{ab}d\zeta^{a}d\zeta^{b} = H_0^{-1/2} (-dt^2 + dx^2) + \frac{6}{R}d\Omega_3^2
\end{align}

\noindent whose scalar curvature is
\begin{align}\label{eq:Rscalar}
   R &= \frac{12}{H_0^{1/2}l^2}\csc^2\psi_0.
\end{align}

At the classical solution, Eq.(\ref{eq:Solution}), Eqs.(\ref{eq:B2tau},\ref{eq:C3tau}) become
\begin{align}
  B_2 &= 0,~~~C_3 = C_3^{(0)} \equiv \frac{3m}{8g_s}\xi(\psi)d\Omega_3.
\end{align}

Plugging all of this into the action, Eq.(\ref{eq:SD4}), and integrating over $x, \chi,\theta,$ and $\phi$, results in an effective action
\begin{align}\label{eq:Lagrangian}
   S^{(b)}_0 &= \int dt \mathcal{L},\\
   \mathcal{L} &= -\alpha N L \sqrt{1 - E^2}\sin^3\psi~ + q L N E \xi(\psi) \nonumber\\
   \alpha &= \frac{l^3}{2\sqrt{2}\pi m\alpha'},~~~q = \frac{3}{2^{3/2} I_0^{1/2}}\alpha
\end{align}

\noindent where $L$ is the periodic length of the probe D4-branes $x$ direction.  Choosing a gauge where $\mathcal{F}_{01} = \dot{A_x}$, leaves us with one conjugate variable with which to perform the Legendre transformation on the Lagrangian Eq.(\ref{eq:Lagrangian})
\begin{align}\label{eq:H}
   H &= \frac{\partial\mathcal{L}}{\partial \dot{A_x}} \dot{A_x} - \mathcal{L}
\end{align}

Because of the periodicity in the field $E$, the conjugate momentum to $A_x$ is quantized to an integer $k$~\cite{Herzog:2002ss}:
\begin{align}
   2\pi\alpha' H_0^{1/2}\frac{\partial\mathcal{L}}{\partial E} = \frac{\partial\mathcal{L}}{\partial \dot{A_x}} = k L.
\end{align}

With this, we find the Hamiltonian, Eq.(\ref{eq:H}), to be
\begin{align}
   H &= \alpha N L \sqrt{\sin^6\psi + \frac{q^2}{\alpha^2}\left(\frac{4 k}{3 N} - \xi\right)^2},
\end{align}

\noindent Minimization with respect to $\psi$ results in the condition
\begin{align}\label{eq:mincon}
  \frac{4 k}{3 N} = \xi(\psi_0) + 3\frac{\alpha^2}{q^2}\sin^2\psi_0\cos\psi_0
\end{align}
\noindent and the minimized Hamiltonian
\begin{align}\label{eq:Hmin}
   H_{min} = T L = \alpha N L \sin^2\psi_0 \sqrt{\sin^2\psi_0 + (3\alpha/q)^2\cos^2\psi_0}\nonumber\\
   \alpha/q \approx 0.3083
\end{align}

\noindent where $T$ is the $k$-string tension.  Note that the parameter $k$ is interpreted as the $k$ quark-anti-quark pairs in $k$-strings.  The tension, Eq.(\ref{eq:Hmin}), and minimization condition, Eq.(\ref{eq:mincon}), form a transcendental equation which can be solved numerically for given $k$ and $N$.

A similar, \emph{back of the envelope}\footnote{We thank A. Armoni for a discussion of this point and its overall relevance.} calculation in the Maldacena-Nastase(M-Na) background~\cite{Maldacena:2001pb}, leads one to a sine law:
\be\label{eq:MNaTension}
T\sim N\sin \frac{\pi k}{N}.
\ee

\noindent Table~\ref{tab:comparetensions} compares the tension calculated from the CGLP background Eq.(\ref{eq:Hmin}) to the sine-law, Eq.(\ref{eq:MNaTension}), Casimir law, and various results from the lattice calculations and Hamiltonian formulation.

\section{Fluctuations of the Classical CGLP Solution}
Employing the same techniques as in~\cite{Zayas:2008hw}, we will now fluctuate around the classical solution and calculate the one loop corrections to the classical energy, Eq.(\ref{eq:Hmin}), $E_1$.  Following~\cite{Zayas:2008hw,pandoz,Bigazzi}, we calculate this correction to be given by the natural log of the functional determinate of the quadratic fluctuation of the partition function, $Z_2$.
\begin{align}
  e^{iE_1T} = Z_2 = \int DX DA D\bar{\Theta} D\Theta e^{i S_2}
\end{align}

\noindent where $S_2$ is the part of the probe D-brane action quadratic in the fluctuations.

To calculate this path integral, one would need to remove the gauge degrees of freedom from symmetries such as $U(1)$ gauge invariance of the gauge fields, diffeomorphism invariance of the probe D-brane, and $\kappa$-symmetry of the supersymmetric D-brane action, via a Fadeev-Popov gauge fixing technique.  We instead use the semi-classical techniques of~\cite{pandoz,Bigazzi,Zayas:2008hw}, and simply solve the gauge fixed equations of motion for the quadratic fluctuations, and sum over the resulting eigenvalues:
\begin{align}\label{eq:oneloopenergy}
   E_1 &= E_1^{(b)} + E_1^{(f)} \\
   \label{eq:fermionandbosononeloopenergy}
   E_1^{(b)} &= \frac{1}{2}\sum \omega_{(b)}~~~E_2^{(f)} = -\frac{1}{2}\sum \omega_{(f)},
\end{align}

\noindent where $E_1^{(b)}$ is the bosonic energy correction and $E_1^{(f)}$ is the fermionic energy correction.

We now seek to the find the equations of motion for the quadratic fluctuations.
The general fluctuations will be given by
\begin{align}\label{eq:generalfluctuation}
   X^{\mu} &= X_0^{\mu} + \delta X^{\mu}~~~\mbox{bosons} \\
   A_a &= A^{(0)}_a + \delta A_a ~~~\mbox{$U(1)$ gauge fields}\\
   \Theta &= 0 + \delta\Theta ~~~\mbox{fermions}
\end{align}

We will first concentrate on the bosonic fluctuations.  For fermionic fluctuations, we will later look to Martucci and collaborators~\cite{m1,m2,m3,m4} to find the proper type IIA supergravity fermionic action.

\subsection{Bosonic Fluctuations}
Of the 10 independent bosonic coordinates, we fluctuate only five of them: the five not statically set to the D4-branes world volume:
\begin{align}
   x^2(\zeta) &= 0 + \delta x^2(\zeta),~~~\psi(\zeta) = \psi_0 + \delta\psi(\zeta),~~~\tau(\zeta) = \tau_0 + \delta\tau(\zeta),~~~\mu^i(\zeta) = \mu^i_0 + \delta\mu^i(\zeta)
\end{align}

\noindent where the subscript zero, as in $\mu^i_0$, refers to the classical value for the field, specified in Eq.(\ref{eq:Solution}).  Under this fluctuation, the induced metric becomes, to quadratic order in the fluctuations
\begin{align}\label{eq:gfluctuated}
   ds^2 &= ds_0^2 + ds_1^2 + ds_2^2 \\
   ds_1^2 &= H_0^{1/2}\frac{l^2}{2}\sin(2\psi_0) d\Omega_3^2 \delta\psi \\
   ds_2^2 &= \left(H_0^{-1/2} \frac{\partial \delta x^2}{\partial\zeta^a}\frac{\partial\delta x^2}{\partial\zeta^b} + H_0^{1/2}l^2\left(\frac{\partial\delta\tau}{\partial\zeta^a}\frac{\partial\delta\tau}{\partial\zeta^b} + \frac{1}{2}\frac{\partial\delta\psi}{\partial\zeta^a}\frac{\partial\delta\psi}{\partial\zeta^b}\right)\right)d\zeta^a d\zeta^b + \nonumber\\
   &~~~+ \delta\tau^2\left[ \frac{H_2}{2H_0^{3/2}}(-dt^2 + dx^2) + H_0^{1/2}l^2\left(A^i_{\alpha} A^i_{\beta} d\Omega_3^{\alpha} d\Omega_3^{\beta} + \sin^2\psi_0(1 - \frac{H_2}{4H_0})d\Omega_3^2\right)\right] + \nonumber\\
          &~~~+ \delta\psi^2 H_0^{1/2}\frac{l^2}{2}\cos(2\psi_0) d\Omega_3^2~~~~~~~~\mbox{$i$ = 1,2},
\end{align}

\noindent $B_2$ becomes
\begin{align}\label{eq:B2fluctuated}
   B_2 &= B_2^{(1)} + B_2^{(2)} \\
   B_2^{(1)} &= \frac{m}{8l}\sin^2\psi_0~\sin\chi~\delta\tau~d\chi \wedge d\theta \\
   B_2^{(2)} &= \frac{m}{8l}\delta\tau~\biggl(\delta\mu^1 \sin^2\psi_0~ \sin^2\chi~\sin\theta~d\theta \wedge d\phi + \delta\mu^2 \sin^2\psi_0~\sin\chi~\sin\theta~d\phi\wedge d\chi + \biggr.\nonumber\\
             &~~~~~~~~~~~~~~~\biggl.+\sin\psi_0~\sin\chi~\sin\theta~\frac{\partial\delta\psi}{\partial\zeta^a}d\zeta^a \wedge d\phi + \delta\psi~\sin(2\psi_0)\sin\chi~d\chi\wedge d\theta\biggr),
\end{align}
\noindent and $C_3$ becomes
\begin{align}\label{eq:C3fluctuated}
  C_3 &= C_3^{(0)} + C_3^{(1)} + C_3^{(2)},\\
  C_3^{(1)} &= \frac{3m}{8 g_s}\delta\psi~\sin^3\psi_0~d\Omega_3,\\
  C_3^{(2)} &= \frac{m}{16 g_s}\left(9 \sin^2\psi_0~\cos\psi_0 \delta\psi^2 - \frac{A^1_{\chi}J^1_{\theta\phi} + A^2_{\theta}J^2_{\phi\chi}}{\sin^2\chi~\sin\theta}\delta\tau^2\right)d\Omega_3
\end{align}

\noindent The fluctuation, Eq.(\ref{eq:generalfluctuation}), leads to a simple expansion for the $U(1)$ gauge field:
\begin{align}\label{eq:Ffluctuated}
   F &= \frac{E}{2\pi\alpha' H_0^{1/2}}dt\wedge dx + \delta F, \nonumber\\
   \delta F &= (\partial_a \delta A_b)d\zeta^a \wedge d\zeta^b.
\end{align}

As the dilaton depends on $\tau$ through Eq.(\ref{eq:dilaton}), the dilaton value, to second order in the tau fluctuations, becomes
\begin{align}\label{eq:dilatonfluctuation}
   e^{-\Phi} &= e^{-\Phi_0}(1 + \frac{H_2}{H_0}\delta\tau^2) \\
   e^{\Phi_0} &= g_s H_0^{1/4}
\end{align}

Using Eq.(\ref{eq:gfluctuated}) through Eq.(\ref{eq:dilatonfluctuation}), we calculate the bosonic action, Eq.(\ref{eq:SD4}), to second order in the fluctuations:
\begin{align}\label{eq:Sbosonic}
  S^{(b)} &= S^{(b)}_0 + S^{(b)}_1 + S^{(b)}_2, \\
  \label{eq:S1bosonic}
  S^{(b)}_1 &= \int \sqrt{-\det(g^{(eff)})}d^5\zeta~k~c_1~\delta F_{tx}, \\
  \label{eq:S2bosonic}
  S^{(b)}_2 &= -\int \sqrt{-\det(g^{(eff)})}d^5\zeta\biggl\{c_x \nabla_a\delta x^2\nabla^a \delta x^2 + c_{\psi} \left[\nabla_a\delta\psi\nabla^a \delta\psi - \frac{R}{2} \delta\psi^2\right] \biggr. + \nonumber\\
            &~~~+ c_{\tau}\left[\nabla_a\delta\tau\nabla^a \delta\tau + m_{\tau}^2(\chi,\theta)\delta\tau^2\right]+ c_A\left[\frac{1}{16 \pi} \delta F^{ab} \delta F_{ab} + j^a \delta A_a \right] + \nonumber\\  &~~~ \biggl. + \mbox{total derivatives} \biggr\},
\end{align}

\noindent where the covariant derivatives are with respect to $g^{(eff)}$, an effective metric on the D4-brane
\begin{align}\label{eq:geff}
   ds^2 &= g^{(eff)}_{ab} d\zeta^a d\zeta^b = \frac{1}{g_{xx}}(-dt^2 + dx^2) + \frac{6}{R}d\Omega_3^2, \nonumber\\
   g_{xx} &= \frac{12^3 A^2 l^6}{H_{min}^2 I_0^2 R^3 m^4},
\end{align}
\noindent where $R$ is the same scalar curvature as in Eq.\ref{eq:Rscalar}. The $U(1)$ gauge current $j^a$, and $m_{\tau}(\chi,\theta)$ are
\begin{align}
   j^a &= \left(-Q_{\psi} \nabla_x \delta\psi,~~~Q_{\psi} \nabla_t \delta\psi, ~~~ Q_{\tau} \frac{\nabla_{\theta}(\sin\theta \delta\tau)}{\sin\chi~\sin\theta},~~~ -Q_{\tau} \frac{\nabla_{\chi}(\sin\chi~\delta\tau)}{\sin^2\chi}, 0\right), \\
   m_{\tau}^2(\chi,\theta) &= m_{\tau 0}^2 + \frac{R}{6}\csc^2\chi \csc^2\theta,
\end{align}

\noindent and the various constants are
\begin{align}
 c_x &= \frac{\mu_4 R^{3/2}l^3 H_{min}}{48 \sqrt{3} g_s A} = \frac{2 l^4}{I_0 m^2}c_{\psi} = \frac{l^4}{I_0 m^2}c_{\tau} = \frac{l^3}{32 I_0^1/2 \pi^3 m\alpha'^2}c_A, \nonumber\\
  Q_{\tau} &= \frac{3 R}{4 I_0^{1/2} \pi^2 12^2 \alpha'},~~~Q_{\psi} = \frac{9 H_{\min}^3 I_0^{9/4} R^{9/2} m^{11/2}}{8\sqrt{6} A^3 \pi^2 l^{15/2} 12^5 \alpha'} \nonumber\\
  m_{\tau 0}^2 &= \frac{l}{16 m I_0^{3/2}}(1 + 8 I_0) + \frac{7}{24}R,~~~c_1 = \left(\frac{H_{min}}{A}\right)^2\left(\frac{R}{12}\right)^{9/2}\frac{I_0^2 m^4}{2 \pi^2 l^3}.
\end{align}

Clearly, $S^{(b)}_1$ is an integral over total derivatives.  This confirms that we are truly fluctuating around a classical solution.  The second order action, $S^{(b)}_2$, as it stands, is difficult to acquire  eigenvalues from, as can be seen in the equations of motion from this action which are listed in Appendix~\ref{app:FullbosonicEquations}.  There we solve one of the equations of motion, and discuss solutions of the others.

To calculate the L\"uscher term, we integrate out the spherical degrees of freedom, $\chi$, $\theta$, and $\phi$, which will leave us with the same number of massless modes as before.  Since the L\"uscher term only depends on massless modes this process should lead us to the same L\"uscher term as would be calculated from the full five dimensional theory.  This was explicitly found to be the case in~\cite{Zayas:2008hw} where the massless modes were independent of the angular degrees of freedom.

To proceed with this integration, we consider the fluctuations to be independent of the $S^3$ variables,
\begin{align}
   \delta X^{\mu} &= \delta X^{\mu}(t,x),~~~\delta A_{a} = \delta A_{a}(t,x)
\end{align}

\noindent and we integrate out the $S^3$ from the action Eq.(\ref{eq:S2bosonic}). This results in an effective action
\begin{align}\label{eq:S2bosoniceff}
   S^{(b)}_{2eff} &= - V_3 \int dt~dx \biggl\{c_x \nabla_m\delta x^2\nabla^m \delta x^2 + c_{\psi} \left[\nabla_m\delta\psi\nabla^m \delta\psi - \frac{R}{2} g_{xx} \delta\psi^2\right] \biggr. + \nonumber\\
            &~~~+ c_{\tau}\left[\nabla_m\delta\tau\nabla^m \delta\tau + m_{\tau e}^2\delta\tau^2\right]+ c_A\left[\frac{1}{g_{xx}16 \pi} \delta F^{mn} \delta F_{mn} + g_{xx} j^m \delta A_m \right] + \nonumber\\
            &~~~ + \frac{c_A}{16 \pi}(\nabla_m \delta A_{\chi}\nabla^m \delta A_{\chi} + 2\nabla_m \delta A_{\theta}\nabla^m \delta A_{\theta} + I_1 \nabla_m \delta A_{\phi}\nabla^m \delta A_{\phi}) + \nonumber\\
            &~~~ + \biggl.\mbox{total derivatives} \biggr\},
\end{align}

\noindent where the indices $m$ and $n$ now sum only over the coordinates $t$ and $x$, and are raised and lowered by the two dimensional Minkowski metric
\begin{align}
   \eta_{mn} d\zeta^m d\zeta^n &= -dt^2 + dx^2,
\end{align}

\noindent and the effective $\delta\tau$ mass, $m_{\tau e}$, the constant $V_3$, and the integral, $I_1$ are
\begin{align}
    m_{\tau e}^2 &= V_3 \left(m_{\tau 0}^2 + \frac{R}{6}I_1\right), ~~~V_3 = 2\pi^2 \left(\frac{R}{6}\right)^{3/2},~~~I_1 = \int_0^{\pi} \csc\theta~d\theta.
\end{align}

The equations of motion of the action, Eq.(\ref{eq:S2bosoniceff}), are
\begin{align}
   &\nabla^2 \delta x^2 = 0 \\
   &\nabla^2 \delta \psi + \frac{R}{2}g_{xx} \delta\psi + \frac{c_A g_{xx} Q_{\psi}}{2 c_{\psi}} \delta F_{tx} = 0 \\
   &\nabla^2\delta\tau - m_{\tau e}^2 \delta \tau = 0 \\
   &\nabla^2 \delta A_i = 0,~~~i = \chi,\theta,\phi \\
   \label{eq:Fmn}
   &\nabla_m \delta F^{mn} = 4\pi g_{xx} j^n
\end{align}

To solve these equations, we move to Fourier space
\begin{align}
  &(\omega^2-p^2)\delta x^2 =0 \\
  &(\omega^2-p^2 + \frac{R}{2}g_{xx}) \delta\psi -i \frac{c_A g_{xx} Q_{\psi}}{2 c_{\psi}} (\omega A_x + p A_t) = 0 \\
   &(\omega^2 - p^2 - m_{\tau e}^2) \delta \tau = 0 \\
   &(\omega^2 - p^2) \delta A_i = 0,~~~i = \chi,\theta,\phi \\
   &p^2\delta A_t + p\omega A_x = -i4\pi g_{xx} Q_{\psi} p \delta\psi \\
   &\omega^2\delta A_x + p\omega A_t = -i4\pi g_{xx} Q_{\psi} \omega \delta\psi,
\end{align}

\noindent and work in temporal gauge, $\delta A_t = 0$.  This leaves us with six eigenvalues
\begin{align}\label{eq:bosonomegas}
  \omega^2 &= \left\{\begin{array}{l}
                 p^2~~~\mbox{4 fold degenerate}  \\
                 p^2 + m_{\tau e}^2 \nonumber\\
                 p^2 + m_{\psi}^2
                 \end{array}
              \right.,
\end{align}
\noindent where

\begin{align}
   m_{\psi}^2 &= \frac{R}{2} - \frac{c_A}{2 c_{\psi}}4\pi Q_{\psi}^2 g_{xx}^2 \\
   45.9992\frac{l}{m} &< m_{\psi}^2 < \infty.
\end{align}

The calculation for the one loop correction to the bosonic $k$-string energy, $E_1^{(b)}$ is given in Appendix~\ref{app:oneloopbosonicEnergy}; the result for large quark separation $L$ is
\begin{align}\label{eq:bosononeloop}
  E_1^{(b)} &= -\frac{\pi}{6 L} - \frac{1}{2}(m_{\tau e} + m_{\psi}).
\end{align}

\noindent This is in contrast to the L\"uscher term found in ~\cite{Luscher:1980fr,Luscher:1980ac,Lucini:2001nv}, which is $-\pi/24 L$.

\subsection{Fermionic Fluctuations}
Following the series of papers~\cite{m1,m2,m3,m4}, we use the $\kappa$-symmetry fixed fermionic action for the probe D4-brane found in~\cite{m4}:
\begin{align}\
   S^{(f)} = \frac{\mu_4}{2}\int d^{5}\zeta e^{-\Phi}\sqrt{-\det M} \bar{\Theta} \Gamma_{D_4}'\biggl[(M^{-1})^{ab}& \Gamma_aD_b^{(0)} +(M^{-1})^{ab}\Gamma_b W_a - \Delta\biggr]\Theta,
\end{align}

\noindent where the definitions are found in Appendix~\ref{app:fermiondefs}.

Applying the fluctuation Eq.(\ref{eq:generalfluctuation}) to the fermions leads us to an action second order in the fermionic fluctuations,
\begin{align}\label{eq:Sffull}
   S^{(f)}_2 = \frac{\mu_4}{2}\int d^{5}\zeta e^{-\Phi}\sqrt{-\det M} \delta\bar{\Theta} \Gamma_{D_4}'\biggl[(M^{-1})^{ab}& \Gamma_aD_b^{(0)} +(M^{-1})^{ab}\Gamma_b W_a - \Delta\biggr]\delta\Theta.
\end{align}

\noindent As we did in the bosonic case, we integrate out the $S^3$, resulting in the effective quadratic fermionic action
\begin{align}\label{eq:fermioneffectiveaction}
   S^{(f)}_{2eff} \propto \int dt dx \delta\bar{\Theta}\Gamma_{D_4}'((M^{-1})^{mn}\Gamma_m\partial_n + M_f) \delta\Theta,~~~ m,n = t,x.
\end{align}

\noindent The matrix $M_f$ is given in Appendix~\ref{app:fermiondefs}.
We solve the Euler equation from this Lagrangian by Fourier transform
\begin{align}\label{eq:DiracFT}
  \Gamma_{D_4}'(i(M^{-1})^{mn}\Gamma_m p_n + M_f) \delta\Theta,~~~ m,n = t,x~~~
  p_t& = -\omega,~~~p_x = p.
\end{align}

\noindent The eigenvalue solutions to this equation, outlined in Appendix~\ref{app:fermiondefs}, are
\begin{align}
  \omega &= \pm\sqrt{p^2 + \alpha_{1} \pm \alpha_{2}}\nonumber\\
  \omega &= \pm\sqrt{p^2 + \alpha_{3} \pm \alpha_{4}}\nonumber\\
   \omega &= \left\{\begin{array}{l}
                    \pm\sqrt{\alpha_{7}(p) + \alpha_{5}(p) \pm \alpha^{+}_{6}(p)} \\
                    \pm\sqrt{\alpha_{7}(p) - \alpha_{5}(p) \pm \alpha^{-}_{6}(p)}
                    \end{array}
                    \right.,
\end{align}

These eigenvalues prove difficult to regulate when one calculates the one loop fermionic energy Eq.(\ref{eq:fermionandbosononeloopenergy}).  This calculation has been postponed until a later time.

\section{Conclusion}
We analyzed the ground state and one loop quantum corrections for $SU(N)$ $k$-strings.  For the ground state, we compared string theory results to lattice gauge theory and Yang-Mills theory results.  We investigated the hypothesis that D4-branes in the CGLP background would describe quarks in the anti-symmetric representation and that D3-branes in the M-Na background would describe quarks in the symmetric representation.  For the tensions, we found, in fact, that both of these were more closely related to anti-symmetric quarks.  We concluded that more research must be done with other probes like D2-branes and D6 in the CGLP background and D5-branes in the M-Na background to find a clearer correlation with symmetric, anti-symmetric, or mixed quark representations.

Furthermore, we analyzed the fluctuations from the ground state for $SU(N)$ $k$-strings in $2+1$ dimensions from a duality relation with a D4-brane probing the type IIA CGLP supergravity background.  We found equations of motion for bosonic and fermionic fluctuations that are relevant for the L\"uscher term.   Interestingly, we found the L\"uscher term was $-\pi/6 L$, the same as in our previous investigation
for 3+1 $k$-strings\footnote{Here we correct a factor of a half missing in~\cite{Zayas:2008hw}.}.  

We predict the L\"uscher term for the $k$-string through the gauge/gravity correspondence.  Since  this result is based on large $N$ with $k/N$ fixed one can only make a qualitative comparison to the work of  ~\cite{Luscher:1980fr,Luscher:1980ac,Lucini:2001nv}.
In fact all that we might be showing in this case is that the quantum excitations of the $k$-strings are different from the lattice result for the {\it fundamental} string.

\section*{Acknowledgments}
The authors would like to gratefully acknowledge conversations with A. Armoni, B. Bringoltz, D. Minic, and  V.P. Nair.
This work is  partially supported by Department of Energy under
grant DE-FG02-95ER40899 to the University of Michigan and the National Science Foundation under award PHY  - 0652983 to the University of Iowa.

\appendix
\section{Five Dimensional Bosonic Equations of Motion for the Fluctuations}
\label{app:FullbosonicEquations}
Applying the variational principle to the action, Eq.(\ref{eq:S2bosonic}), results in the field equations:
\begin{align}\label{eq:x2}
   &\nabla^2\delta x^2 = 0 \\
   \label{eq:psi}
   &\nabla^2\delta\psi + \frac{R}{2}\delta\psi + \frac{c_A Q_{\psi}}{2 c_{\psi}}\delta F_{tx} = 0 \\
   \label{eq:tau}
  &\nabla^2\delta\tau - m_{\tau}^2(\chi,\theta)\delta\tau + \frac{c_A Q_{\tau}}{2 c_{\tau}} \csc\chi \delta F_{\theta\chi} = 0 \\
  \label{eq:FU1}
   &\nabla_{a}\delta F^{ab} - 4 \pi j^{b} = 0,
\end{align}

\noindent where $\nabla_a$ is the covariant derivative compatible with Eq.\ref{eq:geff}.

The solution to Eq.(\ref{eq:x2}) is
\begin{align}\label{eq:x2solution}
   \delta x^2 &= \int d\omega dp \sum_{n \ge l \ge |m|} {\tilde{x}}^{(n,l,m)}(p,\omega) e^{i(p x - \omega t)}Y^{nlm}(\chi,\theta,\phi),
\end{align}

\noindent where the $Y^{nlm}(\chi,\theta,\phi)$ are the spherical harmonics on an $S^3$~\cite{Higuchi:1986wu}
\begin{align}
   Y^{nlm}(\chi,\theta,\phi) &= c_{nl}\frac{1}{\sqrt{\sin\chi}} P^{l+1/2}_{n+1/2}(\cos\chi) Y^{(lm)}(\theta,\phi), \nonumber\\
   c_{nl} &= \sqrt{\frac{(n+1)(n+l+1)!}{(n-l)!}},
\end{align}

\noindent and $P^{l}_{n}(x)$ are the associated Legendre polynomials. The $S^3$ spherical harmonics $Y^{nlm}(\chi,\theta,\phi)$ satisfy the eigenvalue problem
\begin{align}
   \tilde{\nabla}^2 Y^{nlm}(\chi,\theta,\phi) &= -n(n+2)Y^{nlm}(\chi,\theta,\phi),
\end{align}

\noindent where $\tilde{\nabla}^2$ is the Laplacian for an $S^3$ whose action on scalar functions such as $Y^{nlm}(\chi,\theta,\phi)$ is explicitly given by
\begin{align}
   \tilde{\nabla}^2 &= \frac{1}{\sin^2\chi}\left(\partial_{\chi}(\sin^2\chi \partial_{\chi}) + \frac{1}{\sin\theta}\partial_{\theta}(\sin\theta \partial_{\theta}) + \frac{1}{\sin^2\theta}\partial^2_{\phi} \right).
\end{align}

With that said, solving Eq.(\ref{eq:x2}) with the solution Eq.(\ref{eq:x2solution}) results in the eigenvalue problem
\begin{align}
  &\left[\frac{1}{g_{xx}}(\omega^2 - p^2) - \frac{R}{6}n(n+2) \right]\tilde{x} = 0
\end{align}

The rest of the equations prove quite difficult and require perturbation theory to solve; their solution is not given here.

\section{Calculation of the One Loop Bosonic Energy}
\label{app:oneloopbosonicEnergy}
Following~\cite{Zayas:2008hw,pandoz,Bigazzi}, we define the one loop bosonic energy as a sum over the bosonic frequencies, Eq.(\ref{eq:bosonomegas}),
\begin{align}
   E^{(b)}_1 \equiv \frac{1}{2} \sum_{p}\left( 4 p + \sqrt{p^2 + m_{\tau e}^2} + \sqrt{p^2 + m_{\psi}^2} \right).
\end{align}

\noindent To describe quark-anti-quark sources affixed to the end of the string, we use vanishing boundary conditions
\begin{align}
   p &= n \pi/L
\end{align}

\noindent As in~\cite{Zayas:2008hw}, we use $\zeta$-function regularization to regularize the massless modes
\begin{align}
   \sum_{n = 1}^{\infty} n ~&\longrightarrow ~ \zeta(-1) = -\frac{\pi}{12}.
\end{align}

For the massive modes, instead of using the techniques of~\cite{Bertoldi:2004rn}, we regulate the sum by using a cutoff
\begin{align}
   \sum_{n=1}^{\infty} \sqrt{n^2 + M^2} ~ &\longrightarrow ~\sum_{n=1}^{[M]} \sqrt{n^2 + M^2}
\end{align}

\noindent where $[M]$ is the largest integer less than $M$.  Now we are able to use the binomial expansion for the square root
\begin{align}
   \sum_{n=1}^{[M]} \sqrt{n^2 + M^2} &= M \sum_{n=1}^{[M]} \sum_{q=0}^{\infty}\binom{\frac{1}{2}}{q} \left(\frac{n}{M}\right)^{2q} \nonumber\\
       &= M \sum_{q=0}^{\infty}\binom{\frac{1}{2}}{q}M^{-2q} \sum_{n=1}^{[M]}n^{2q} \nonumber\\
       &\approx M \sum_{q=0}^{\infty}\binom{\frac{1}{2}}{q}M^{-2q} \sum_{n=1}^{\infty} n^{2q} \nonumber\\
       & ~ \longrightarrow ~ M \sum_{q=0}^{\infty}\binom{\frac{1}{2}}{q}M^{-2q}\delta^0_q \zeta(0) \nonumber\\
       &= -\frac{M}{2}
\end{align}

\noindent where we have again used $\zeta$-function regularization, and the approximation $[M] \to \infty$ is valid for large $M \propto  L$, large $L$ corresponding to large quark separation.

Using these results, we find for large quark separation $L$, the one loop bosonic energy is
\begin{align}
   E^{(b)}_1 &= -\frac{\pi}{6 L} -\frac{1}{2}(m_{\tau e} + m_{\psi})\tag{\ref{eq:bosononeloop}}.
\end{align}
\section{Fermionic Dp-brane Action}
\label{app:fermiondefs}
Throughout this paper, and unless otherwise noted, Latin indices $a,b,c,\dots$ are D-brane indices, with the exception of $i,j$, and $k$, which will \emph{always} refer to the unit $S^2$ coordinates, $\mu^i$.  Greek indices, $\alpha,\beta,\mu,\nu,\dots$ are 10 dimensional curved indices, and underlined Greek indices, $\underline{\mu},\underline{\nu},\underline{\alpha},\dots$ are 10 dimensional flat indices.

The fermionic action for the quadratic fluctuations was found in~\cite{m1,m2,m3,m4}.  It is given by
\begin{align}
   S^{(f)} = \frac{\mu_4}{2}\int d^{5}\zeta e^{-\Phi}\sqrt{-\det M} \delta\bar{\Theta} \Gamma_{D_4}'\biggl[(M^{-1})^{ab}& \Gamma_aD_b^{(0)} +(M^{-1})^{ab}\Gamma_b W_a - \Delta\biggr]\delta\Theta. \tag{\ref{eq:Sffull}}
\end{align}

\noindent where $\delta\Theta$ is a 32 component spinor, constrained by
\begin{align}\label{eq:Thetaconstraint}
   \Gamma^{11} \delta\Theta = \delta\Theta.
\end{align}

\noindent where $\Gamma^{11} = \Gamma^{\underline{0123456789}}$, and  $\Gamma^{\underline{\mu_1\mu_2\dots\mu_n}}$ is the totally antisymmetric product of gamma matrices.  The flat gamma matrices satisfy a Clifford algebra
\begin{align}
   \{ \Gamma^{\underline{\mu}},\Gamma^{\underline{\nu}} \} = 2 \eta^{\underline{\mu\nu}}
\end{align},

\noindent where $\eta^{\underline{\mu\nu}}$ is the 10 dimensional Minkowski metric.  A useful consequence of this anti-commutation relation is the following identity
\begin{align}
    \Gamma_{\mu\nu\alpha\beta} = \Gamma_{\mu}\Gamma_{\nu}\Gamma_{\alpha}\Gamma_{\beta}, ~~~\mu \ne \nu \ne \alpha \ne \beta,
\end{align}
\noindent which can be generalized to any number of gamma matrices.

We define $\Delta = \Delta^{(1)} + \Delta^{(2)}$, $M = g + \mathcal{F}$, $\mathcal{F}_4 = F_4 + H_3 \wedge C_1$, $F_2 = dC_1$, where
\begin{align}\label{eq:Da0}
   D_a^{(0)} &= \partial_a + \frac{1}{4}\Omega_a^{\underline{\mu\nu}}\Gamma_{\underline{\mu\nu}}+ \frac{1}{4 \cdot 2!}H_{a\mu\nu}\Gamma^{\mu\nu} \\
   W_a &= -\frac{1}{8}e^{\Phi}\left(\frac{1}{2}F_{\mu\nu}\Gamma^{\mu\nu} + \frac{1}{4!}\mathcal{F}_{\mu\nu\alpha\beta}\Gamma^{\mu\nu\alpha\beta}\right)\Gamma_a \\
   \Delta^{(1)} &= \frac{1}{2}(\Gamma^{\mu}\partial_{\mu}\Phi + \frac{1}{2\cdot 3!} H_{\mu\alpha\beta}\Gamma^{\mu\alpha\beta}) \\
   \Delta^{(2)} &= \frac{1}{8} e^{\Phi}\left(\frac{3}{2!}F_{\mu\nu}\Gamma^{\mu\nu} - \frac{1}{4!}\mathcal{F}_{\mu\nu\alpha\beta}\Gamma^{\mu\nu\alpha\beta}\right)\\
   \Gamma_{D_p}' &=1 - \frac{\sqrt{-\det g}}{\sqrt{-\det M_0}} \Gamma_{D_p}^{(0)}(\Gamma^{11})^{p/2+1}\sum_{q \ge 0} \frac{(-1)^q (\Gamma^{11})^q}{q! 2^q}\Gamma^{a_1 a_2\dots a_{2q}}\mathcal{F}_{a_1 a_2} \dots \mathcal{F}_{a_{2q-1}a_{2q}} \\
   \label{eq:Dp0}
   \Gamma_{D_p}^{(0)} &= \frac{\epsilon^{a_1 a_2 \dots a_{p+1}}}{(p+1)!\sqrt{-\det g}}\Gamma_{a_1 a_2 \dots a_{p+1}}
\end{align}

\noindent and $\epsilon^{a_1 a_2 \cdots a_{p+1}}$ is a density, i.e., takes values of $\pm 1$, or 0.

The 10 dimensional curved $\Gamma_{\mu}$'s are related to the 10 dimensional flat $\Gamma_{\underline{\mu}}$'s by the frame fields, $\Gamma_{\mu} = e^{\underline{\mu}}_{~\mu}\Gamma_{\underline{\mu}}$, and the D-brane $\Gamma_a$'s are pulled back from the curved 10 dimensional $\Gamma_{\mu}$'s:  $\Gamma_a = \frac{\partial X^{\mu}}{\partial \zeta^a} \Gamma_{\mu}$.

The frame-fields,
\begin{align}
  G_{\mu\nu} &= e^{\underline{\mu}}_{~\mu} e^{\underline{\nu}}_{~\nu}\eta_{\underline{\mu\nu}}
\end{align}
\noindent for the CGLP background can be written in a 10 dimensional representation as
\begin{align}
   e^{\underline{0}}_{~0} &= e^{\underline{1}}_{~1} = e^{\underline{2}}_{~2} = H^{-1/4},~~~e^{\underline{9}}_{~9} = l f H^{1/4},   \nonumber\\
   e^{\underline{3}}_{~3} &= \csc\theta e^{\underline{4}}_{~4} = \csc\chi\csc\theta e^{\underline{5}}_{~5} = \csc\psi\csc\chi\csc\theta e^{\underline{6}}_{~6} = lbH^{1/4}, \nonumber\\
   e^{\underline{7}}_{~\mu} &= lH^{1/4}~a \frac{\partial{\mu^i}}{\partial\tilde{\theta}} A^j_{\alpha} \epsilon^{ijk}\mu^k,~~~e^{\underline{8}}_{~\mu} = lH^{1/4}a \csc\tilde{\theta}\frac{\partial{\mu^i}}{\partial\tilde{\phi}} A^j_{\alpha} \epsilon^{ijk}\mu^k,~~~\alpha = 4,5,6
\end{align}

\noindent with parametrization of the unit $S^2$, $(\mu^i)^2 = 1$, given by
\begin{align}
   \mu^1 &= \sin\tilde{\theta}~\cos\tilde{\phi},~~~\mu^2 = \sin\tilde{\theta}~\sin\tilde{\phi},~~~\mu^3 = \cos\tilde{\theta}
\end{align}

\noindent In the above, the the 10 independent bosonic coordinates are numbered $0\dots9$ as
\begin{align}
   X^{\mu} = (t,x,x^2,\psi,\chi,\theta,\phi,\tilde{\theta},\tilde{\phi},\tau)
\end{align}

We use the frame fields to calculate the spin connection for the 10-dimensional bosonic space
\begin{align}\label{eq:10dSpinConnection}
   \Omega_{\mu}^{~\underline{\alpha\beta}} = \Omega_{\mu}^{~\underline{\beta\alpha}} &= \eta^{\underline{\beta\rho}}e^{\underline{\alpha}}_{~\alpha}\left(\partial_{\mu}e_{\underline{\rho}}^{~\alpha} + e_{\underline{\rho}}^{~\nu}\Gamma^{\alpha}_{~\mu\nu}\right)
\end{align}

\noindent where the Christoffels, $\Gamma^{\alpha}_{~\mu\nu}$, and the inverse frame-fields, $e_{\underline{\mu}}^{~\mu}$, are given by
\begin{align}
   \Gamma^{\alpha}_{~\mu\nu} &= \frac{1}{2}G^{\alpha\beta}\left(\partial_{\mu}G_{\beta\nu} + \partial_{\nu}G_{\beta\mu} - \partial_{\beta}G_{\mu\nu}\right)\nonumber\\
   e_{\underline{\mu}}^{~\mu} &= \eta_{\underline{\mu\nu}}G^{\mu\nu}e^{\underline{\nu}}_{~\nu}
\end{align}

\noindent With this, we calculate the spin connection for the CGLP background.  Pulling its lowered index back to the D4-brane and evaluating it at the classical solution, we find the only non-vanishing components to be
\begin{align}
  &\Omega_{\chi}^{~\underline{98}} = \Omega_{\chi}^{~\underline{43}} = \cos\psi_0, &&\Omega_{\theta}^{~\underline{53}} = \Omega_{\theta}^{~\underline{79}} = \cos\psi_0~\sin\chi, \nonumber\\
  &\Omega_{\theta}^{~\underline{54}} = \Omega_{\theta}^{~\underline{87}} = \cos\chi, &&\Omega_{\phi}^{~\underline{63}} = \Omega_{\phi}^{~\underline{87}} = \cos\psi_0~\sin\chi~\sin\theta, \nonumber\\
  &\Omega_{\phi}^{~\underline{64}} = \Omega_{\phi}^{~\underline{97}} = \cos\chi~\sin\theta,
  &&\Omega_{\phi}^{~\underline{65}} = \Omega_{\phi}^{~\underline{98}} = \cos\theta,
\end{align}

\noindent With this we calculate the term in the action which contains the spin connection to be
\begin{align}
   \frac{1}{4}\left(M^{-1}\right)^{ab}\Gamma_a \Omega_b^{~\underline{\mu\nu}}\Gamma_{\underline{\mu\nu}} &= M_c + \cot\chi~M_1 + \csc\chi~\cot\theta~M_2 \nonumber\\
   M_c &= \frac{1}{2}\sqrt{\frac{R}{6}}\cos\psi_0(3\Gamma_{\underline{3}} + \Gamma_{\underline{498}}+\Gamma_{\underline{579}}+\Gamma_{\underline{687}}) \nonumber\\
   M_1 &= \frac{1}{2}\sqrt{\frac{R}{6}}(2\Gamma_{\underline{4}} + \Gamma_{\underline{587}} + \Gamma_{\underline{697}}) \nonumber\\
   M_2 &= \frac{1}{2}\sqrt{\frac{R}{6}}(\Gamma_{\underline{5}} + \Gamma_{\underline{698}})
\end{align}

\noindent This is the only term in the action that has $\theta,\chi$ dependence, modulo the measure.  Many of the formulas in Eq.(\ref{eq:Da0})-Eq.(\ref{eq:Dp0}) simplify
\begin{align}
   W_a &= -\frac{1}{8}e^{\Phi_0}\frac{1}{4!}F_{\mu\nu\alpha\beta}\Gamma^{\mu\nu\alpha\beta}\Gamma_a \\
   \Delta^{(1)} &= \frac{1}{4!}H_{\alpha\mu\nu}\Gamma^{\alpha\mu\nu},~~~\Delta^{(2)} = -\frac{1}{8 \cdot 4!}e^{\Phi_0} F_{\alpha\beta\mu\nu}\Gamma^{\alpha\beta\mu\nu}\\
   \Gamma_{D_4}' &= 1 - \frac{\epsilon^{abcde}\Gamma_{abcde}}{5!\sqrt{-\det M}}\Gamma^{11}(1 - \frac{1}{2}\Gamma^{11}\Gamma^{ab}\mathcal{F}_{ab}),
\end{align}

\noindent all of which are, again, $\chi,\theta$ independent.

As in the bosonic case, we investigate $S^3$ independent solutions for the quadratic fluctuations
\begin{align}
   \delta\Theta = \delta\Theta(t,x)
\end{align}

\noindent leaving us with an action of the form
\begin{align}
   S^{(f)}_{2eff} \propto \int dt dx\int d\chi d\theta d\phi &\sin^2\chi~\sin\theta \delta\bar{\Theta}\Gamma_{D_4}'((M^{-1})^{mn}\Gamma_m\partial_n + M_f + \nonumber\\
         & + \cot\chi~M_1 + \csc\chi~\cot\theta~M_2) \delta\Theta,~~~ m,n = t,x,
\end{align}

\noindent where
\begin{align}
  M_f &= M_c + \left(M^{-1}\right)^{ab}\left(\frac{1}{8}\Gamma_a H_{b\mu\nu}\Gamma^{\mu\nu} + \Gamma_b W_a\right) - \Delta
\end{align}

Integrating out the $S^3$ as before, it is easy to see that the terms proportional to $M_1$ and $M_2$ integrate to zero, leaving us with
\begin{align}
   S^{(f)}_{2eff} \propto \int dt dx \delta\bar{\Theta}\Gamma_{D_4}'((M^{-1})^{mn}\Gamma_m\partial_n + M_f) \delta\Theta,~~~ m,n = t,x. \tag{\ref{eq:fermioneffectiveaction}}
\end{align}

We solve the Euler equation from this action by Fourier transform
\begin{align}
  \Gamma_{D_4}'(i(M^{-1})^{mn}\Gamma_m p_n + M_f) \delta\Theta,~~~ m,n = t,x~~~
  p_t& = -\omega,~~~p_x = p.\tag{\ref{eq:DiracFT}}
\end{align}

We now pick a representation for the $32 \times 32$ gamma matrices
\begin{align}\label{eq:flatgamma}
  \Gamma^{\underline{0}} &= i \sigma^1 \otimes \sigma^1 \otimes \sigma^1 \otimes \sigma^3 \otimes \sigma^0,~~~\Gamma^{\underline{1}} = \sigma^1 \otimes \sigma^1 \otimes \sigma^2 \otimes \sigma^0 \otimes \sigma^0 \nonumber\\
  \Gamma^{\underline{2}} &= \sigma^1 \otimes \sigma^1 \otimes \sigma^1 \otimes \sigma^1 \otimes \sigma^3,~~~\Gamma^{\underline{3}} = \sigma^1 \otimes \sigma^1 \otimes \sigma^1 \otimes \sigma^1 \otimes \sigma^2 \nonumber\\
  \Gamma^{\underline{4}} &= \sigma^1 \otimes \sigma^1 \otimes \sigma^3 \otimes \sigma^0 \otimes \sigma^0,~~~\Gamma^{\underline{5}} = \sigma^1 \otimes \sigma^2 \otimes \sigma^0 \otimes \sigma^0 \otimes \sigma^0 \nonumber\\
  \Gamma^{\underline{6}} &= -\sigma^1 \otimes \sigma^1 \otimes \sigma^1 \otimes \sigma^1 \otimes \sigma^1,~~~\Gamma^{\underline{7}} = \sigma^1 \otimes \sigma^3 \otimes \sigma^0 \otimes \sigma^0 \otimes \sigma^0 \nonumber\\
  \Gamma^{\underline{8}} &= \sigma^2 \otimes \sigma^0 \otimes \sigma^0 \otimes \sigma^0 \otimes \sigma^0,~~~\Gamma^{\underline{9}} = \sigma^1 \otimes \sigma^1 \otimes \sigma^1 \otimes \sigma^2 \otimes \sigma^4
\end{align}

\noindent where $\otimes$ means tensor product, and the $\sigma^{\mu}$ are the Pauli spin matrices, augmented with the identity:
\begin{align}\label{eq:Paulispin}
   \sigma^0 &= \left(\begin{array}{l l}
                     1 & 0 \\
                     0 & 1
                     \end{array}
                \right),~~~
   \sigma^1 &= \left(\begin{array}{l l}
                       0 & 1 \\
                       1 & 0
                      \end{array}
                \right),~~~
   \sigma^2 &= \left(\begin{array}{l l}
                      0 & -i \\
                      i & 0
                     \end{array}
               \right),~~~
   \sigma^3 &= \left(\begin{array}{l l}
                      1 & 0 \\
                      0 & -1
                     \end{array}
               \right)
\end{align}

This representation leaves us with a diagonal $\Gamma^11$, and so through the constraint, Eq.(\ref{eq:Thetaconstraint}), we are able to set the lower 16 components of $\delta\Theta$ to zero.  At the same time, this reduces the 32 equations in Eq.(\ref{eq:DiracFT}) to 16 independent equations.  These equations can be reorganized into the following form
\begin{align}
  \omega \delta\Theta &= H_f \delta\Theta
\end{align}

\noindent where the Hamiltonian has the block diagonal form
\begin{align}
   H_f = \left(\begin{array}{lll}
                      H_1 & 0 & 0 \\
                      0 & H_2 & 0 \\
                      0 & 0 & H_3
               \end{array}
         \right),
\end{align}

\noindent and $H_1$ and $H_2$ are $4 \times 4$ matrices, and $H_3$ is an $8 \times 8$ matrix
\begin{align}\label{eq:Hi}
   H_1 &= \left(\begin{array}{llll}
                    p & 0 & -c_i & c_a \\
                    0 & p & c_b & -c_i \\
                    -c_j & c_c & -p & 0 \\
                    c_d & -c_j & 0 & -p
                \end{array}
          \right),
   H_2 = \left(\begin{array}{llll}
                    p & 0 & c_i & c_e \\
                    0 & p & c_f & c_i \\
                    c_j & c_g & -p & 0 \\
                    c_h & c_j & 0 & -p
                \end{array}
          \right), \nonumber\\
   H_3 &= \left(\begin{array}{llllllll}
                  -p & 0 & c_j & -c_c & 0 & 0 & -c_k & 0 \\
                  0 & -p & -c_d & c_j & 0 & 0 & 0 & -c_k \\
                  c_i & -c_a & p & 0 & -c_n & 0 & 0 & 0 \\
                  -c_b & c_i & 0 & p & 0 & -c_n & 0 & 0 \\
                  0 & 0 & -c_k & 0 & -p & 0 & -c_j & -c_g \\
                  0 & 0 & 0 & -c_k & 0 & -p & -c_h & -c_j \\
                  -c_n & 0 & 0 & 0 -c_i & -c_e & p & 0 \\
                  0 & -c_n & 0 & 0 & -c_f & -c_i & 0 & p
                \end{array}
          \right)
\end{align}

\noindent where the $c's$ are constants.

The eigenvalues of $H_f$ are
\begin{align}
  \omega &= \pm\sqrt{p^2 + \alpha_{1} \pm \alpha_{2}} \nonumber\\
  \omega &= \pm\sqrt{p^2 + \alpha_{3} \pm \alpha_{4}} \nonumber\\
   \omega &= \left\{\begin{array}{l}
                    \pm\sqrt{\alpha_{7}(p) + \alpha_{5}(p) \pm \alpha^{+}_{6}(p)} \\
                    \pm\sqrt{\alpha_{7}(p) - \alpha_{5}(p) \pm \alpha^{-}_{6}(p)}
                    \end{array}
                    \right.,
\end{align}

where $\alpha_1, \alpha_2, \alpha_3,$ and $\alpha_4$ are constants combinations of the $c's$ in~\ref{eq:Hi},  and $\alpha_{5}, \alpha_6^{\pm},$ and $\alpha_7$ are functions of p:

\begin{align}
    \alpha_{7}(p) &= \alpha_{7}^{(0)}+\alpha_{7}^{(2)}p^2,\nonumber\\
    \alpha_{5}^2 &= \sum_{n = 0,2,4}\alpha_{5}^{(n)}p^n + 4\beta_{1}(p),\nonumber\\
    (\alpha_{6}^\pm)^2 &= 2\alpha_{5}^{2}(p) - 3\beta_{1}(p) \pm \beta_{5}(p),\nonumber\\
    \beta_{1}(p) &= \frac{1}{12}\left(\frac{\beta_{3}(p)}{\beta_{2}(p)}+\beta_{2}(p)\right),\nonumber\\
    2\beta_{2}^3(p) &= \beta_{4}(p) +\sqrt{\beta_{4}^3(p) - 4\beta_{3}^3(p)},\nonumber\\
    \beta_{3}(p) &= \sum_{n = 0,2,,8}\beta_{3}^{(n)}p^n,\nonumber\\
    \beta_{4}(p) &= \sum_{n = 0,2,,12}\beta_{4}^{(n)}p^n,\nonumber\\
    \beta_{5}(p) &= \alpha_{5}^{-1}\sum_{n = 0,2,,6}\beta_{5}^{(n)}p^n.
\end{align}

\noindent Here, the $\alpha_{i}^{(n)}$ and $\beta_{i}^{(n)}$ are constant combinations of the $c's$ from~\ref{eq:Hi}.
\bibliographystyle{utphys}
\bibliography{bibliography}

\providecommand{\href}[2]{#2}\begingroup\raggedright\begin{thebibliography}{10}

\bibitem{Polyakov:1986cs}
A.~M. Polyakov, ``{Fine Structure of Strings},''
\href{http://dx.doi.org/10.1016/0550-3213(86)90162-8}{{\em Nucl. Phys.} {\bf
  B268} (1986)  406--412}.

\bibitem{Karabali:1997wk}
D.~Karabali, C.-j. Kim, and V.~P. Nair, ``{Planar Yang-Mills theory:
  Hamiltonian, regulators and mass gap},''
  \href{http://dx.doi.org/10.1016/S0550-3213(98)00309-5}{{\em Nucl. Phys.} {\bf
  B524} (1998)  661--694},
\href{http://arxiv.org/abs/hep-th/9705087}{{\tt arXiv:hep-th/9705087}}.

\bibitem{Karabali:2000gy}
D.~Karabali, C.-j. Kim, and V.~P. Nair, ``{Manifest covariance and the
  Hamiltonian approach to mass gap in (2+1)-dimensional Yang-Mills theory},''
  \href{http://dx.doi.org/10.1103/PhysRevD.64.025011}{{\em Phys. Rev.} {\bf
  D64} (2001)  025011},
\href{http://arxiv.org/abs/hep-th/0007188}{{\tt arXiv:hep-th/0007188}}.

\bibitem{Cvetic:2001ma}
M.~Cvetic, G.~W. Gibbons, H.~Lu, and C.~N. Pope, ``{Supersymmetric non-singular
  fractional D2-branes and NS-NS 2-branes},''
  \href{http://dx.doi.org/10.1016/S0550-3213(01)00236-X}{{\em Nucl. Phys.} {\bf
  B606} (2001)  18--44},
\href{http://arxiv.org/abs/hep-th/0101096}{{\tt arXiv:hep-th/0101096}}.

\bibitem{Shifman:2005eb}
M.~Shifman, ``{k strings from various perspectives: QCD, lattices, string
  theory and toy models},'' {\em Acta Phys. Polon.} {\bf B36} (2005)
  3805--3836,
\href{http://arxiv.org/abs/hep-ph/0510098}{{\tt arXiv:hep-ph/0510098}}.

\bibitem{hk}
C.~P. Herzog and I.~R. Klebanov, ``{On string tensions in supersymmetric SU(M)
  gauge theory},'' \href{http://dx.doi.org/10.1016/S0370-2693(02)01155-3}{{\em
  Phys. Lett.} {\bf B526} (2002)  388--392},
\href{http://arxiv.org/abs/hep-th/0111078}{{\tt arXiv:hep-th/0111078}}.

\bibitem{Herzog:2002ss}
C.~P. Herzog, ``{String tensions and three dimensional confining gauge
  theories},'' \href{http://dx.doi.org/10.1103/PhysRevD.66.065009}{{\em Phys.
  Rev.} {\bf D66} (2002)  065009},
\href{http://arxiv.org/abs/hep-th/0205064}{{\tt arXiv:hep-th/0205064}}.

\bibitem{Ridgway:2007vh}
J.~M. Ridgway, ``{Confining k-string tensions with D-branes in super Yang-
  Mills theories},''
  \href{http://dx.doi.org/10.1016/j.physletb.2007.02.054}{{\em Phys. Lett.}
  {\bf B648} (2007)  76--83},
\href{http://arxiv.org/abs/hep-th/0701079}{{\tt arXiv:hep-th/0701079}}.

\bibitem{tye}
H.~Firouzjahi, L.~Leblond, and S.~H. Henry~Tye, ``{The (p,q) string tension in
  a warped deformed conifold},'' {\em JHEP} {\bf 05} (2006)  047,
\href{http://arxiv.org/abs/hep-th/0603161}{{\tt arXiv:hep-th/0603161}}.

\bibitem{Zayas:2008hw}
L.~A. Pando~Zayas, V.~G.~J. Rodgers, and K.~Stiffler, ``{Term for k-string
  Potential from Holographic One Loop Corrections},''
  \href{http://dx.doi.org/10.1088/1126-6708/2008/12/036}{{\em JHEP} {\bf 12}
  (2008)  036},
\href{http://arxiv.org/abs/0809.4119}{{\tt arXiv:0809.4119 [hep-th]}}.

\bibitem{Ambjorn:1984mb}
J.~Ambjorn, P.~Olesen, and C.~Peterson, ``{Stochastic Confinement and
  Dimensional Reduction. 1. Four- Dimensional SU(2) Lattice Gauge Theory},''
\href{http://dx.doi.org/10.1016/0550-3213(84)90475-9}{{\em Nucl. Phys.} {\bf
  B240} (1984)  189}.

\bibitem{Ambjorn:1984yu}
J.~Ambjorn, P.~Olesen, and C.~Peterson, ``{THREE-DIMENSIONAL LATTICE GAUGE
  THEORY AND STRINGS},''
\href{http://dx.doi.org/10.1016/0550-3213(84)90193-7}{{\em Nucl. Phys.} {\bf
  B244} (1984)  262}.

\bibitem{Ambjorn:1984me}
J.~Ambjorn, P.~Olesen, and C.~Peterson, ``{OBSERVATION OF A STRING IN
  THREE-DIMENSIONAL SU(2) LATTICE GAUGE THEORY},''
\href{http://dx.doi.org/10.1016/0370-2693(84)91352-2}{{\em Phys. Lett.} {\bf
  B142} (1984)  410}.

\bibitem{Karabali:1998yq}
D.~Karabali, C.-j. Kim, and V.~P. Nair, ``{On the vacuum wave function and
  string tension of Yang- Mills theories in (2+1) dimensions},''
  \href{http://dx.doi.org/10.1016/S0370-2693(98)00751-5}{{\em Phys. Lett.} {\bf
  B434} (1998)  103--109},
\href{http://arxiv.org/abs/hep-th/9804132}{{\tt arXiv:hep-th/9804132}}.

\bibitem{Teper:1998te}
M.~J. Teper, ``{SU(N) gauge theories in 2+1 dimensions},''
  \href{http://dx.doi.org/10.1103/PhysRevD.59.014512}{{\em Phys. Rev.} {\bf
  D59} (1999)  014512},
\href{http://arxiv.org/abs/hep-lat/9804008}{{\tt arXiv:hep-lat/9804008}}.

\bibitem{Bringoltz:2006zg}
B.~Bringoltz and M.~Teper, ``{A precise calculation of the fundamental string
  tension in SU(N) gauge theories in 2+1 dimensions},''
  \href{http://dx.doi.org/10.1016/j.physletb.2006.12.056}{{\em Phys. Lett.}
  {\bf B645} (2007)  383--388},
\href{http://arxiv.org/abs/hep-th/0611286}{{\tt arXiv:hep-th/0611286}}.

\bibitem{Bringoltz:2008nd}
B.~Bringoltz and M.~Teper, ``{Closed k-strings in SU(N) gauge theories : 2+1
  dimensions},'' \href{http://dx.doi.org/10.1016/j.physletb.2008.04.052}{{\em
  Phys. Lett.} {\bf B663} (2008)  429--437},
\href{http://arxiv.org/abs/0802.1490}{{\tt arXiv:0802.1490 [hep-lat]}}.

\bibitem{Armoni:2003ji}
A.~Armoni and M.~Shifman, ``{On k-string tensions and domain walls in N = 1
  gluodynamics},'' \href{http://dx.doi.org/10.1016/S0550-3213(03)00409-7}{{\em
  Nucl. Phys.} {\bf B664} (2003)  233--246},
\href{http://arxiv.org/abs/hep-th/0304127}{{\tt arXiv:hep-th/0304127}}.

\bibitem{Armoni:2003nz}
A.~Armoni and M.~Shifman, ``{Remarks on stable and quasi-stable k-strings at
  large N},'' \href{http://dx.doi.org/10.1016/j.nuclphysb.2003.08.021}{{\em
  Nucl. Phys.} {\bf B671} (2003)  67--94},
\href{http://arxiv.org/abs/hep-th/0307020}{{\tt arXiv:hep-th/0307020}}.

\bibitem{Luscher:1980fr}
M.~Luscher, K.~Symanzik, and P.~Weisz, ``{Anomalies of the Free Loop Wave
  Equation in the WKB Approximation},''
\href{http://dx.doi.org/10.1016/0550-3213(80)90009-7}{{\em Nucl. Phys.} {\bf
  B173} (1980)  365}.

\bibitem{Luscher:1980ac}
M.~Luscher, ``{Symmetry Breaking Aspects of the Roughening Transition in Gauge
  Theories},''
\href{http://dx.doi.org/10.1016/0550-3213(81)90423-5}{{\em Nucl. Phys.} {\bf
  B180} (1981)  317}.

\bibitem{Lucini:2001nv}
B.~Lucini and M.~Teper, ``{Confining strings in SU(N) gauge theories},''
  \href{http://dx.doi.org/10.1103/PhysRevD.64.105019}{{\em Phys. Rev.} {\bf
  D64} (2001)  105019},
\href{http://arxiv.org/abs/hep-lat/0107007}{{\tt arXiv:hep-lat/0107007}}.

\bibitem{Douglas:1995nw}
M.~R. Douglas and S.~H. Shenker, ``{Dynamics of SU(N) supersymmetric gauge
  theory},'' \href{http://dx.doi.org/10.1016/0550-3213(95)00258-T}{{\em Nucl.
  Phys.} {\bf B447} (1995)  271--296},
\href{http://arxiv.org/abs/hep-th/9503163}{{\tt arXiv:hep-th/9503163}}.

\bibitem{Seiberg:1994aj}
N.~Seiberg and E.~Witten, ``{Monopoles, duality and chiral symmetry breaking in
  N=2 supersymmetric QCD},''
  \href{http://dx.doi.org/10.1016/0550-3213(94)90214-3}{{\em Nucl. Phys.} {\bf
  B431} (1994)  484--550},
\href{http://arxiv.org/abs/hep-th/9408099}{{\tt arXiv:hep-th/9408099}}.

\bibitem{Hanany:1997hr}
A.~Hanany, M.~J. Strassler, and A.~Zaffaroni, ``{Confinement and strings in
  M{QCD}},'' \href{http://dx.doi.org/10.1016/S0550-3213(97)00651-2}{{\em Nucl.
  Phys.} {\bf B513} (1998)  87--118},
\href{http://arxiv.org/abs/hep-th/9707244}{{\tt arXiv:hep-th/9707244}}.

\bibitem{Lucini:2004my}
B.~Lucini, M.~Teper, and U.~Wenger, ``{Glueballs and k-strings in SU(N) gauge
  theories: Calculations with improved operators},'' {\em JHEP} {\bf 06} (2004)
   012,
\href{http://arxiv.org/abs/hep-lat/0404008}{{\tt arXiv:hep-lat/0404008}}.

\bibitem{DelDebbio:2001kz}
L.~Del~Debbio, H.~Panagopoulos, P.~Rossi, and E.~Vicari, ``{k-string tensions
  in SU(N) gauge theories},''
  \href{http://dx.doi.org/10.1103/PhysRevD.65.021501}{{\em Phys. Rev.} {\bf
  D65} (2002)  021501},
\href{http://arxiv.org/abs/hep-th/0106185}{{\tt arXiv:hep-th/0106185}}.

\bibitem{Karabali:2009rg}
D.~Karabali, V.~P. Nair, and A.~Yelnikov, ``{The Hamiltonian Approach to
  Yang-Mills (2+1): An Expansion Scheme and Corrections to String Tension},''
\href{http://arxiv.org/abs/0906.0783}{{\tt arXiv:0906.0783 [hep-th]}}.

\bibitem{Gomis:2006sb}
J.~Gomis and F.~Passerini, ``{Holographic Wilson loops},'' {\em JHEP} {\bf 08}
  (2006)  074,
\href{http://arxiv.org/abs/hep-th/0604007}{{\tt arXiv:hep-th/0604007}}.

\bibitem{Gomis:2006im}
J.~Gomis and F.~Passerini, ``{Wilson loops as D3-branes},'' {\em JHEP} {\bf 01}
  (2007)  097,
\href{http://arxiv.org/abs/hep-th/0612022}{{\tt arXiv:hep-th/0612022}}.

\bibitem{Maldacena:2001pb}
J.~M. Maldacena and H.~S. Nastase, ``{The supergravity dual of a theory with
  dynamical supersymmetry breaking},'' {\em JHEP} {\bf 09} (2001)  024,
\href{http://arxiv.org/abs/hep-th/0105049}{{\tt arXiv:hep-th/0105049}}.

\bibitem{Karabali:2007mr}
D.~Karabali and V.~P. Nair, ``{The robustness of the vacuum wave function and
  other matters for Yang-Mills theory},''
  \href{http://dx.doi.org/10.1103/PhysRevD.77.025014}{{\em Phys. Rev.} {\bf
  D77} (2008)  025014},
\href{http://arxiv.org/abs/0705.2898}{{\tt arXiv:0705.2898 [hep-th]}}.

\bibitem{Luscher:2004ib}
M.~Luscher and P.~Weisz, ``{String excitation energies in SU(N) gauge theories
  beyond the free-string approximation},'' {\em JHEP} {\bf 07} (2004)  014,
\href{http://arxiv.org/abs/hep-th/0406205}{{\tt arXiv:hep-th/0406205}}.

\bibitem{Aharony:2009gg}
O.~Aharony and E.~Karzbrun, ``{On the effective action of confining strings},''
\href{http://arxiv.org/abs/0903.1927}{{\tt arXiv:0903.1927 [hep-th]}}.

\bibitem{Luscher:2002qv}
M.~Luscher and P.~Weisz, ``{Quark confinement and the bosonic string},'' {\em
  JHEP} {\bf 07} (2002)  049,
\href{http://arxiv.org/abs/hep-lat/0207003}{{\tt arXiv:hep-lat/0207003}}.

\bibitem{ks}
I.~R. Klebanov and M.~J. Strassler, ``{Supergravity and a confining gauge
  theory: Duality cascades and chiSB-resolution of naked singularities},'' {\em
  JHEP} {\bf 08} (2000)  052,
\href{http://arxiv.org/abs/hep-th/0007191}{{\tt arXiv:hep-th/0007191}}.

\bibitem{Maldacena:2000yy}
J.~M. Maldacena and C.~Nunez, ``{Towards the large N limit of pure N = 1 super
  Yang Mills},'' \href{http://dx.doi.org/10.1103/PhysRevLett.86.588}{{\em Phys.
  Rev. Lett.} {\bf 86} (2001)  588--591},
\href{http://arxiv.org/abs/hep-th/0008001}{{\tt arXiv:hep-th/0008001}}.

\bibitem{Cvetic:2000mh}
M.~Cvetic, H.~Lu, and C.~N. Pope, ``{Brane resolution through transgression},''
  \href{http://dx.doi.org/10.1016/S0550-3213(01)00050-5}{{\em Nucl. Phys.} {\bf
  B600} (2001)  103--132},
\href{http://arxiv.org/abs/hep-th/0011023}{{\tt arXiv:hep-th/0011023}}.

\bibitem{candelas}
P.~Candelas and X.~C. de~la Ossa, ``{Comments on Conifolds},''
\href{http://dx.doi.org/10.1016/0550-3213(90)90577-Z}{{\em Nucl. Phys.} {\bf
  B342} (1990)  246--268}.

\bibitem{pandoz}
L.~A. Pando~Zayas, J.~Sonnenschein, and D.~Vaman, ``{Regge trajectories
  revisited in the gauge / string correspondence},''
  \href{http://dx.doi.org/10.1016/j.nuclphysb.2003.12.006}{{\em Nucl. Phys.}
  {\bf B682} (2004)  3--44},
\href{http://arxiv.org/abs/hep-th/0311190}{{\tt arXiv:hep-th/0311190}}.

\bibitem{Bigazzi}
F.~Bigazzi, A.~L. Cotrone, L.~Martucci, and L.~A. Pando~Zayas, ``{Wilson loop,
  Regge trajectory and hadron masses in a Yang- Mills theory from semiclassical
  strings},'' \href{http://dx.doi.org/10.1103/PhysRevD.71.066002}{{\em Phys.
  Rev.} {\bf D71} (2005)  066002},
\href{http://arxiv.org/abs/hep-th/0409205}{{\tt arXiv:hep-th/0409205}}.

\bibitem{m1}
D.~Marolf, L.~Martucci, and P.~J. Silva, ``{Fermions, T-duality and effective
  actions for D-branes in bosonic backgrounds},'' {\em JHEP} {\bf 04} (2003)
  051,
\href{http://arxiv.org/abs/hep-th/0303209}{{\tt arXiv:hep-th/0303209}}.

\bibitem{m2}
D.~Marolf, L.~Martucci, and P.~J. Silva, ``{Actions and fermionic symmetries
  for D-branes in bosonic backgrounds},'' {\em JHEP} {\bf 07} (2003)  019,
\href{http://arxiv.org/abs/hep-th/0306066}{{\tt arXiv:hep-th/0306066}}.

\bibitem{m3}
D.~Marolf, L.~Martucci, and P.~J. Silva, ``{The explicit form of the effective
  action for F1 and D- branes},'' {\em Class. Quant. Grav.} {\bf 21} (2004)
  S1385--S1390,
\href{http://arxiv.org/abs/hep-th/0404197}{{\tt arXiv:hep-th/0404197}}.

\bibitem{m4}
L.~Martucci, J.~Rosseel, D.~Van~den Bleeken, and A.~Van~Proeyen, ``{Dirac
  actions for D-branes on backgrounds with fluxes},''
  \href{http://dx.doi.org/10.1088/0264-9381/22/13/014}{{\em Class. Quant.
  Grav.} {\bf 22} (2005)  2745--2764},
\href{http://arxiv.org/abs/hep-th/0504041}{{\tt arXiv:hep-th/0504041}}.

\bibitem{Higuchi:1986wu}
A.~Higuchi, ``{SYMMETRIC TENSOR SPHERICAL HARMONICS ON THE N SPHERE AND THEIR
  APPLICATION TO THE DE SITTER GROUP SO(N,1)},''
\href{http://dx.doi.org/10.1063/1.527513}{{\em J. Math. Phys.} {\bf 28} (1987)
  1553}.

\bibitem{Bertoldi:2004rn}
G.~Bertoldi, F.~Bigazzi, A.~L. Cotrone, C.~Nunez, and L.~A. Pando~Zayas, ``{On
  the universality class of certain string theory hadrons},''
  \href{http://dx.doi.org/10.1016/j.nuclphysb.2004.08.044}{{\em Nucl. Phys.}
  {\bf B700} (2004)  89--139},
\href{http://arxiv.org/abs/hep-th/0401031}{{\tt arXiv:hep-th/0401031}}.

\end{thebibliography}\endgroup

\end{document}